\documentclass[final,3p,times]{elsarticle}

\usepackage{amssymb}

\usepackage{amsmath}
\usepackage{isotope}
\usepackage{dcolumn}
\usepackage{bm}
\usepackage{float}
\usepackage{subcaption}
\usepackage{hyperref}

\journal{Nuclear Physics A}

\begin{document}

\begin{frontmatter}



\title{An investigation of radiative proton--capture reactions in the Cd--In mass region}


\author[inst1]{P.~Vasileiou\corref{tim}}
\cortext[tim]{Corresponding author: polvasil@phys.uoa.gr}
\author[inst1]{T.J.~Mertzimekis}
\author[inst1]{A.~Chalil\fnref{saclay}}
\author[inst1]{C.~Fakiola}
\author[inst1]{I~Karakasis}
\author[inst1]{A.~Kotsovolou}
\author[inst1]{S.~Pelonis}
\author[inst1]{A.~Zyriliou}

\affiliation[inst1]{organization={National \& Kapodistrian University of Athens},
            addressline={Zografou Campus}, 
            city={Athens},
            postcode={GR-15784}, 
            country={Greece}}

\author[inst2]{M.~Axiotis}
\author[inst2]{A.~Lagoyannis}

\affiliation[inst2]{organization={Institute of Nuclear and Particle Physics, NCSR "Demokritos"},
            city={Aghia Paraskevi},
            postcode={GR-15310}, 
            country={Greece}}
\fntext[saclay]{Present address: IRFU, CEA, Université Paris-Saclay, France}

\begin{abstract}
The reaction network in the neutron--deficient part of the nuclear chart around
$A \sim 100$ contains several nuclei of importance to astrophysical processes,
such as the {\em p}--process. This work reports on the results from recent 
experimental studies of the radiative proton-capture reactions
\isotope[112,114]{Cd}$(p,\gamma)$\isotope[113,115]{In}. Experimental cross
sections for the reactions have been measured for proton beam energies residing
inside the respective Gamow windows for each reaction, using isotopically
enriched \isotope[112]{Cd} and \isotope[114]{Cd} targets. Two different
techniques, the in--beam $\gamma$--ray spectroscopy and the activation method
have been employed, with the latter considered necessary to account for the
presence of low--lying isomers in \isotope[113]{In} ($E_{\gamma} \approx 392$~keV,
$t_{1/2} \approx 100$~min), and \isotope[115]{In} ($E_{\gamma} \approx 336$~keV,
$t_{1/2} \approx 4.5$~h). Following the measurement of the total reaction cross
sections, the astrophysical {\em S} factors have been additionally deduced.
The experimental results are compared with 
Hauser--Feshbach theoretical calculations carried out with the most recent
version of {\sc TALYS}. The results are discussed in terms of their significance
to the various parameters entering the models.

\end{abstract}




\end{frontmatter}


\section{Introduction}

The neutron--deficient stable nuclides with mass $A\geq 74$ are bypassed by the {\em s--}
and {\em r--}process nucleosynthetic mechanisms\footnote{{\em s} stands for slow and {\em r}
for rapid neutron capture process, respectively.}. This mass region, ranging
from \isotope[74]{Se} to \isotope[196]{Hg} contains about 35 such species of nuclei, commonly
referred to as {\em p--nuclei}, with the letter {\em p} corresponding to their lower
neutron--to--proton ratio $(N/Z)$, relative to other stable isotopes of the same element.
The origin of this particular group of nuclei, which is the rarest among the stable species,
with solar abundances typically of a factor of $\sim 10^2$ times lower compared to the adjacent
{\em s} and {\em r} nuclides~\cite{arnould,iliadis}, has been a long standing puzzle in nuclear
astrophysics~\cite{B2HF,AGW}.

It is generally accepted that the more abundant {\em s} and {\em r} nuclei serve as seeds for
the {\em p}--process, leading to the production of neutron-deficient nuclei through a network
of $(p,\gamma)$ radiative proton--capture reactions, $\beta$ decays, electron captures (EC),
$(\gamma,n)$, $(\gamma,p)$ and $(\gamma,\alpha)$ photodisintegrations. The {\em p}--process
mechanism dominated by photodisintegrations is often referred to as $\gamma$ process~\cite{woosley1}.

The {\em p}--process is assumed to take place in different zones inside a core collapse
supernova, placing the peak temperature for this process in the range
$T_{peak}\sim 2$--$3$~GK~\cite{iliadis}. A single-degenerate type-Ia supernova scenario is
also suitable for the {\em p}--process to occur~\cite{travaglio}. Additional contributions
to the production of {\em p}--nuclei arise from several explosive nucleosynthesis scenarios,
such as
the {\em rp}--process~\cite{rp1_schatz,rp2_wanajo},
the {\em np}--process~\cite{pn_goriely}, and
the {\em $\nu$p}--process~\cite{nup_pinedo}.

The {\em p}--process spans among roughly 2\,000 nuclei, forming a vast reaction network of
about 20\,000 reactions~\cite{arnould,iliadis}. Thus, within that framework, the majority
of the reaction rates need to be estimated, a task often performed by means of the
Hauser--Feshbach statistical model~\cite{HF}.

In the effort to constrain the model parameters, experimental input is invaluable.
Among others, cross--section measurements of radiative proton--capture reactions
can have an essential contribution towards understanding the $\gamma$-- and
{\em p}--process (and other reaction mechanisms, such as the {\em r} or {\em s} process~\cite{Rauscher,Wisshak,Gyurky}), serving to constrain the various model
parameters, thus improving predictions for currently unmeasured reactions, while
enabling the calculation of various important photodisintegration 
constants~\cite{iliadis2}. Radiative proton-capture reactions can offer data
to determine reaction rates in areas of the nuclear chart, which may
cross the pathways of the {\em s} and {\em r} processes. As an example,
the odd-even \isotope[113]{In} nucleus, which is studied in the present work,
is nowadays widely accepted that it is not a ``pure'' {\em p} nucleus, but has non--negligible contributions from the {\em s} and {\em r} processes (see
discussions in refs.~\cite{Psaltis,Pignatari}).

Recent works by our group focusing on \isotope[107,109]{Ag}~\cite{Khaliel} and
\isotope[112]{Cd}~\cite{Psaltis,Khaliel_hnps,Zyriliou_hnps} proton--capture reactions have
provided experimental input in this mass regime for both ground and isomeric states
populated via the in--beam~\cite{Rolfs1973} and activation methods~\cite{activ1}. Precise
measurements of cross sections in reactions of Ag, Cd and In isotopes, as well as other
neighboring nuclei, around mass 100--120, using charged particle or photon probes, are necessary for
reducing the experimental uncertainties and constraining the various pathways
of {\em p}, {\em r} and {\em s} processes in this area of the nuclear chart~\cite{Rauscher,Wisshak,Gyurky}.

\isotope[110]{Cd}, which is populated in the \isotope[109]{Ag}$(p,\gamma)$\isotope[110]{Cd}
reaction, is considered an important {\em p}--nucleus~\cite{Rauscher}, while the role of
\isotope[113]{In} populated in the \isotope[112]{Cd}$(p,\gamma)$\isotope[113]{In} reaction
is important for understanding any potential linking of the {\em r}-- and {\em s}--processes
with the reactions rate affecting the {\em p}--process in this mass regime~\cite{Dillmann}.
In addition, the slightly heavier \isotope[114]{Cd} nucleus is involved in the
{\em s}--process~\cite{Wisshak} and to the best of our knowledge, no experimental data of
proton--capture total cross sections are available at low energies (around 3~MeV) for this
nucleus~\cite{EXFOR}.

The present work focuses on radiative proton--capture reactions in the Cd/In region, through
measurements of the proton--induced reaction cross sections in \isotope[112,114]{Cd} isotopes at
astrophysically relevant energies; that is, energies lying within the corresponding Gamow
windows for the particular reactions ($\sim 1.6-4.8$~MeV). Earlier work of our group had focused
on the reaction \isotope[112]{Cd}$(p,\gamma)$\isotope[113]{In} at proton beam energies below
the $(p,n)$ energy threshold~\cite{Psaltis}. This work extends the $(p,\gamma)$ cross--section
measurements to the energy region above the $(p,n)$ reaction threshold, while still inside
the Gamow window, for the first time, thus enabling the extraction of important information
regarding the optical model potentials (OMP), nuclear level densities (NLD) and gamma strength
functions ($\gamma$SF) entering the theoretical models.

An additional focus is the study of the radiative proton--capture reaction \isotope[114]{Cd}$(p,\gamma)$\isotope[115]{In}, including the known isomeric state at 336~keV.
For this case, an older set of experimental cross sections is reported in Ref.~\cite{Cd7},
while in a more recent work~\cite{Cd4} the focus has been on the production of the unstable
isotope \isotope[115m]{In} for medical applications. In both cases, the isomeric transition
in \isotope[115]{In} was examined, exclusively. The present results on \isotope[115]{In} are
compared with the overall scarce experimental data existing in the energy range of
interest~\cite{Cd4,Cd7}. Furthermore, cross--section measurements in the $(p,n)$ reaction
channel are reported, extending the experimental information to lower energies than those
existing in literature~\cite{Cd1,Cd2,Cd3,Cd4,Cd5,Cd6,Cd9}.

Overall, the experimental cross section results presented in this work can serve as constraints to
model parameters entering the theoretical calculations. The predictions of the Hauser--Feshbach
statistical model have been deduced using the latest version (v1.95) of {\sc TALYS}
code~\cite{talys}, in a systematic way, exploring the sensitivity on various parameters
entering the theoretical calculations. After performing calculations with all possible
OMP+NLD+$\gamma$SF model combinations, using the default input parameters in {\sc TALYS},
``best-fit'' choices are presented for comparison with the experimental data.

\label{sec:intro}

\section{Experimental Details}

Measurements for the study of the radiative proton-capture reactions in
\isotope[112,114]{Cd} were carried out at the 5.5~MV T11 Tandem Van de
Graaff accelerator of NCSR ``Demokritos'', Greece. Both
the in-beam~\cite{Psaltis,Khaliel} and the activation~\cite{Psaltis,activ1}
methods were employed to account for the
presence of low-lying isomeric states in the populated nuclei \isotope[113,115]{In}.

\subsection{The reactions with proton beams}
\label{ssec:2.1}

The reaction \isotope[112]{Cd}($p,\gamma$)\isotope[113]{In} 
($Q$=6081.2(2)~keV) and the \isotope[114]{Cd}$(p,\gamma)$\isotope[115]{In}
($Q$=6810.4(3)~keV)~\cite{nndc} were both studied at three proton laboratory
energies in total, 3.4, 3.5 and 3.6~MeV, and 3.0, 3.5 and 4.0~MeV, respectively.
All energies lie inside the corresponding Gamow windows for the
reactions ($\sim1.6$--$4.8$~MeV, in both cases), calculated for the temperature range $T_9\sim 1.7$--$3.3$~GK~\cite{iliadis}. During the experiments the targets were irradiated with proton beam currents ranging from 60--90~enA, which were kept at a steady value throughout each beam energy run, so as to minimize uncertainties resulting from current fluctuations.

\subsection{The targets}
\label{ssec:tgt}

\subsubsection{The \isotope[112]{Cd} target}
\label{112tgt}

A multi-layer target was irradiated during the experiments comprising a front
layer of $99.7$\% enriched \isotope[112]{Cd} evaporated on a \isotope[nat]{Bi}
layer, backed by an \isotope[nat]{In} layer, and a thick \isotope[nat]{Cu} layer.
The thickness of the \isotope[112]{Cd} layer had been earlier measured~\cite{Psaltis}
and found equal to $\delta_{avg}=0.99(5)$~mg/cm$^2$. The target was mounted at the
center of the scattering chamber and was oriented at a $30^{\circ}$ angle with respect
to the beam direction, to prevent any masking effects from the aluminum  target holder
that could have resulted in attenuation of gamma rays recorded in the surrounding
high-purity Germanium (HPGe) detectors, particularly, the one positioned at $90^{\circ}$
(see also Refs.~\cite{Psaltis,Khaliel}). The resulted effective thickness of the
$^{112}\mathrm{Cd}$ target due to this rotation, was 
$\delta=\delta_{avg}/\cos{30^{\circ}}=1.14(6)$~mg/cm$^2$.

Proton-beam energy losses inside the target were calculated using \textsc{SRIM}2013~\cite{SRIM},
and were found to be in the range $\Delta E=54$--$52$~keV for the proton-beam energies
of 3.4--3.6~MeV, in the laboratory frame, respectively. Assuming that the reactions occur in the middle
of the \isotope[112]{Cd} layer, the effective energy in the lab frame is found as
(see also Table~\ref{tab:112}):
\begin{equation}
    E_{eff}=E_p-\frac{\Delta E}{2} \label{eq:en_loss}
\end{equation}

\subsubsection{The \isotope[114]{Cd} target}
\label{114tgt}

The \isotope[114]{Cd} target used in the experiments comprised of a thin front
layer of 99.7\% enriched \isotope[114]{Cd} evaporated on a \isotope[nat]{Ta}
layer. The use of an enriched target, as in the case of \isotope[112]{Cd}, was imperative
due to the very low cross sections of the reaction.

\begin{figure}[ht]
\centering
\includegraphics[width=0.5\textwidth]{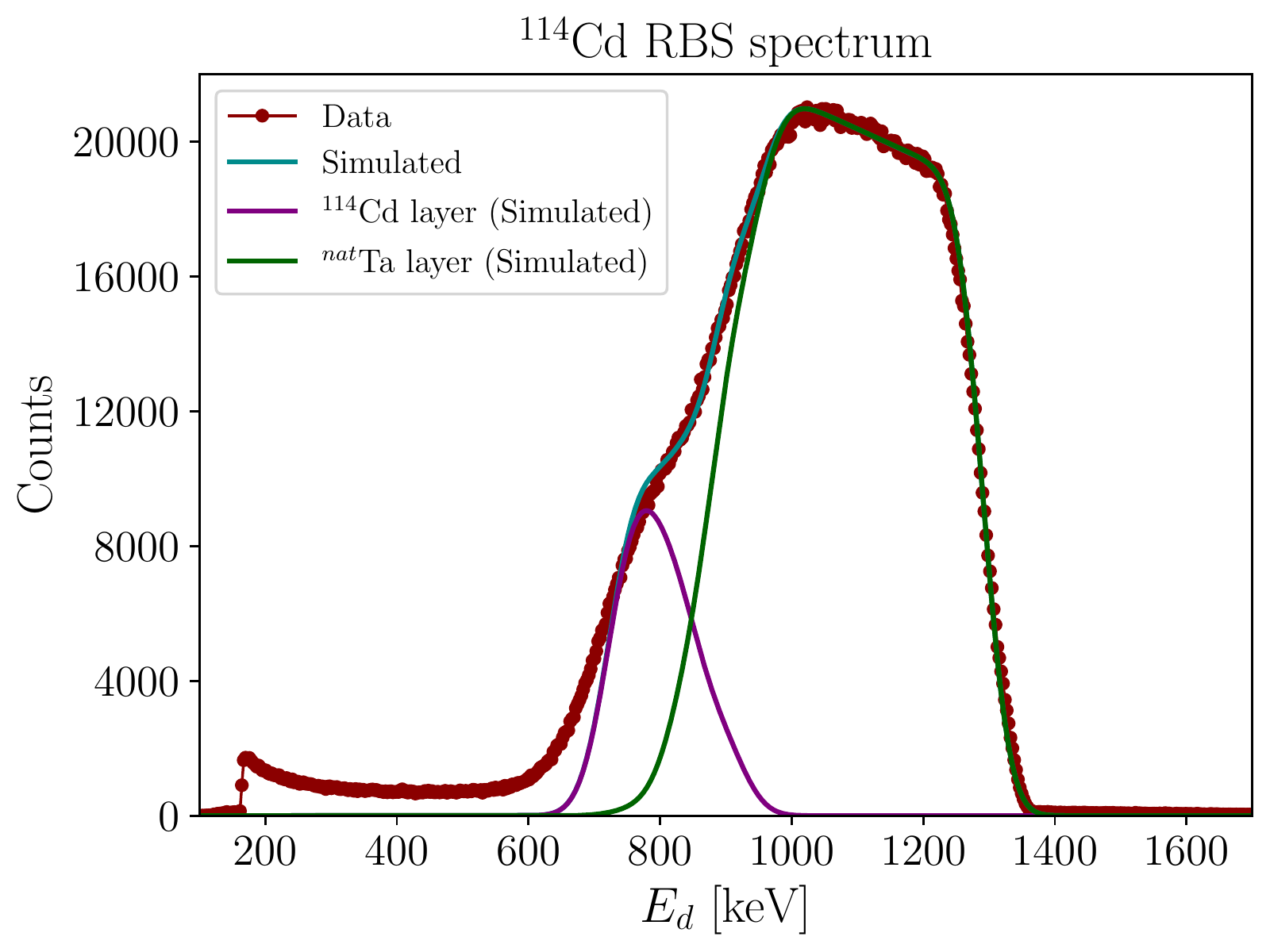}
\caption{\label{fig:RBS}%
The RBS spectrum of the \isotope[114]{Cd} target, produced through bombardment of the target with a deuteron beam of $E_d=1.35$~MeV. The simulation performed in order to obtain the target thickness was carried out with the \textsc{SIMNRA} code~\cite{simnra}.}
\end{figure}

The \isotope[114]{Cd} layer thickness was measured via the Rutherford Backscattering technique,
resulting in a thickness value of $\delta_{RBS}=0.482(13)$~mg/cm$^2$ (see Fig.~\ref{fig:RBS}).
For the same reasons described in~\ref{112tgt}, the target was rotated inside
the chamber at $30^{\circ}$ with respect to the beam, resulting in an effective
thickness of $\delta=0.557(16)$~mg/cm$^2$.

Proton beam energy losses were found to be $\Delta E= 28$--$24$~keV, for beam
energies in the range 3.0--4.0~MeV in the laboratory frame.

\subsection{Detection apparatus and experimental methods}
\label{ssec:exp_methods}

\begin{figure}[ht]
\centering
\includegraphics[width=0.5\columnwidth]{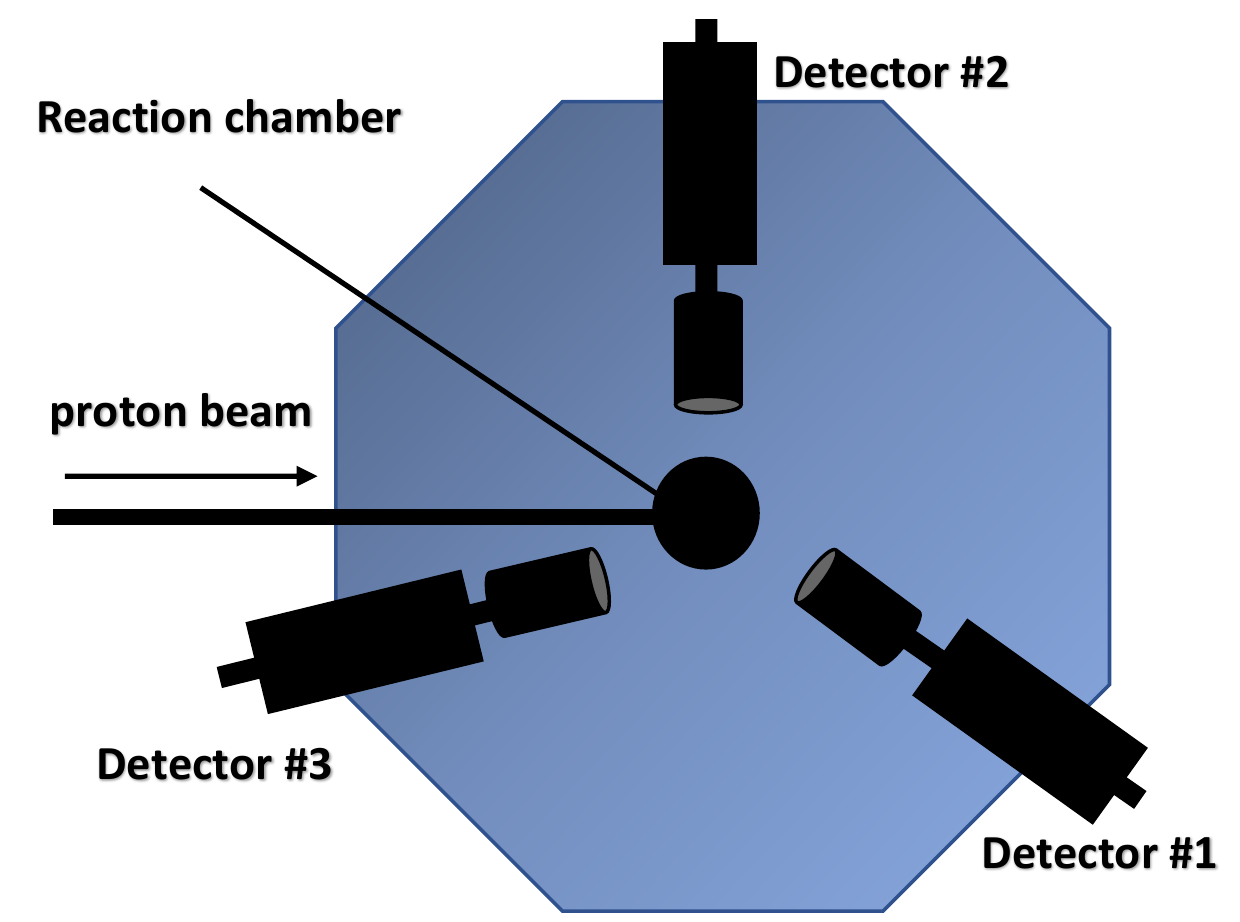}
\caption{\label{fig:det_setup}%
Schematic representation of the experimental setup. The target chamber was surrounded by an array of three HPGe detectors placed on a turntable to measure $\gamma$ singles at three different angles.}
\end{figure}

\begin{figure}[ht]
\centering
\includegraphics[width=0.5\columnwidth]{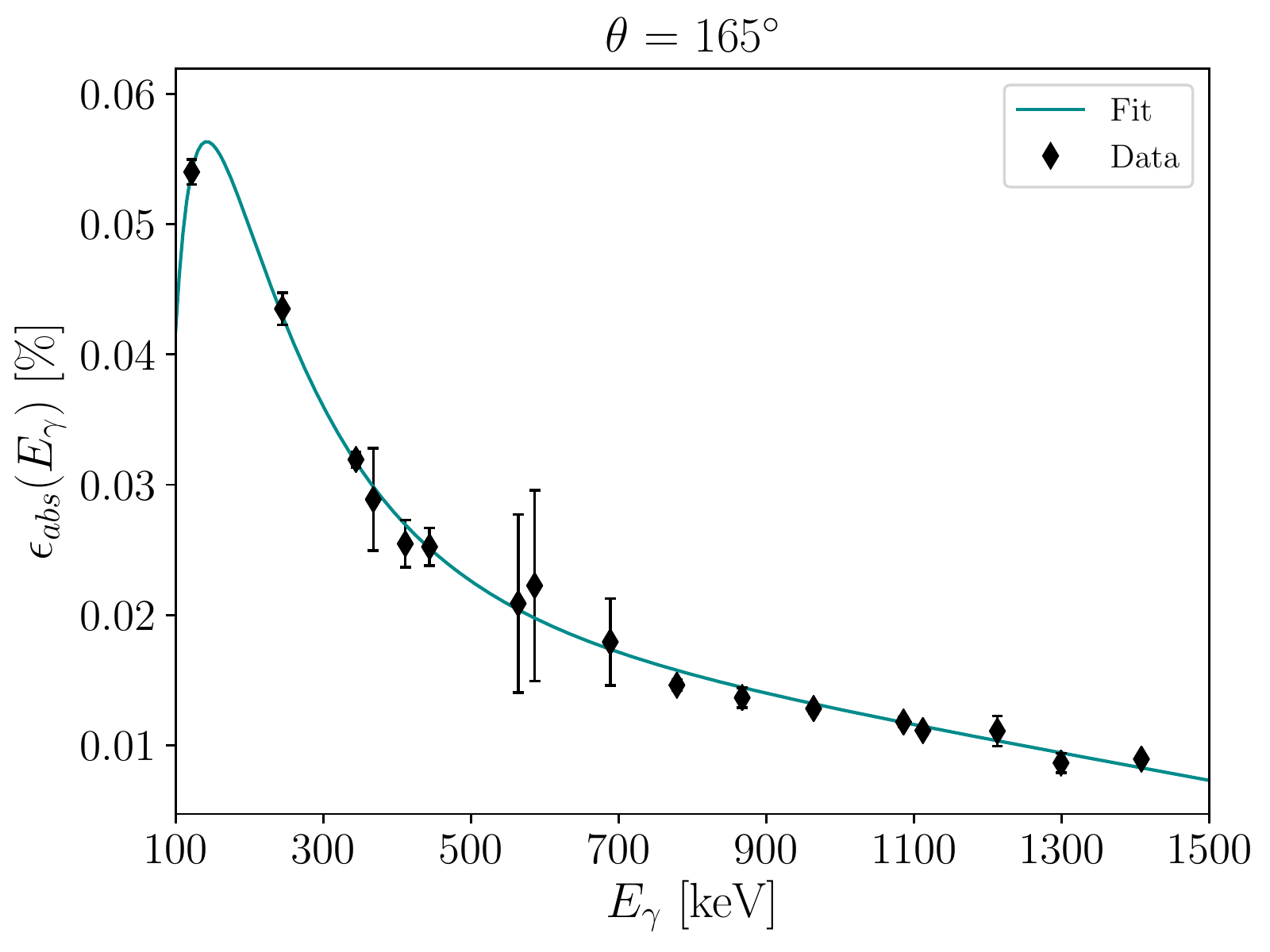}
\caption{\label{fig:det_eff}%
Typical absolute efficiency curve for the detectors employed in the experiments. The particular one corresponds to the detector placed at $165^{\circ}$.}
\end{figure}

An array of three HPGe detectors were mounted on an octagonal turntable having a maximum
radius of 2.4~m (see sketch in Fig.~\ref{fig:det_setup}). Detectors 1, 2 and 3 were positioned
at $55^{\circ}$, $90^{\circ}$ and $165^{\circ}$, respectively, with reference to the beam
direction. Their distances from the target were 34.5, 27.7 and 33.0~cm, respectively. Energy
calibrations and absolute efficiency measurements (Fig.~\ref{fig:det_eff}) were performed
using a standard \isotope[152]{Eu} point source, placed at the exact target position.
The nuclear electronics setup described in Ref.~\cite{Khaliel} was employed in order
to record the 8k--channel spectra in singles mode (i.e. without the use of $\gamma$--particle or $\gamma$--$\gamma$ coincidence techniques).

\section{Data Analysis and Results}
\label{data_analysis}

Due to the structural properties of the nuclei \isotope[113,115]{In}, two
different methods were employed for the study of the cross sections of the
radiative proton-capture reactions: in-beam $\gamma$-ray spectroscopy to study
prompt de--excitations, and the activation technique to study longer--lived
isomeric states populated in the reactions.

\subsection{The reaction \isotope[112]{Cd}$(p,\gamma)$\isotope[113]{In}}
\label{ssec:112Cd}

\subsubsection{In-beam measurements}
\label{112ib}

\begin{figure}[ht]
\centering
\includegraphics[width=0.9\textwidth]{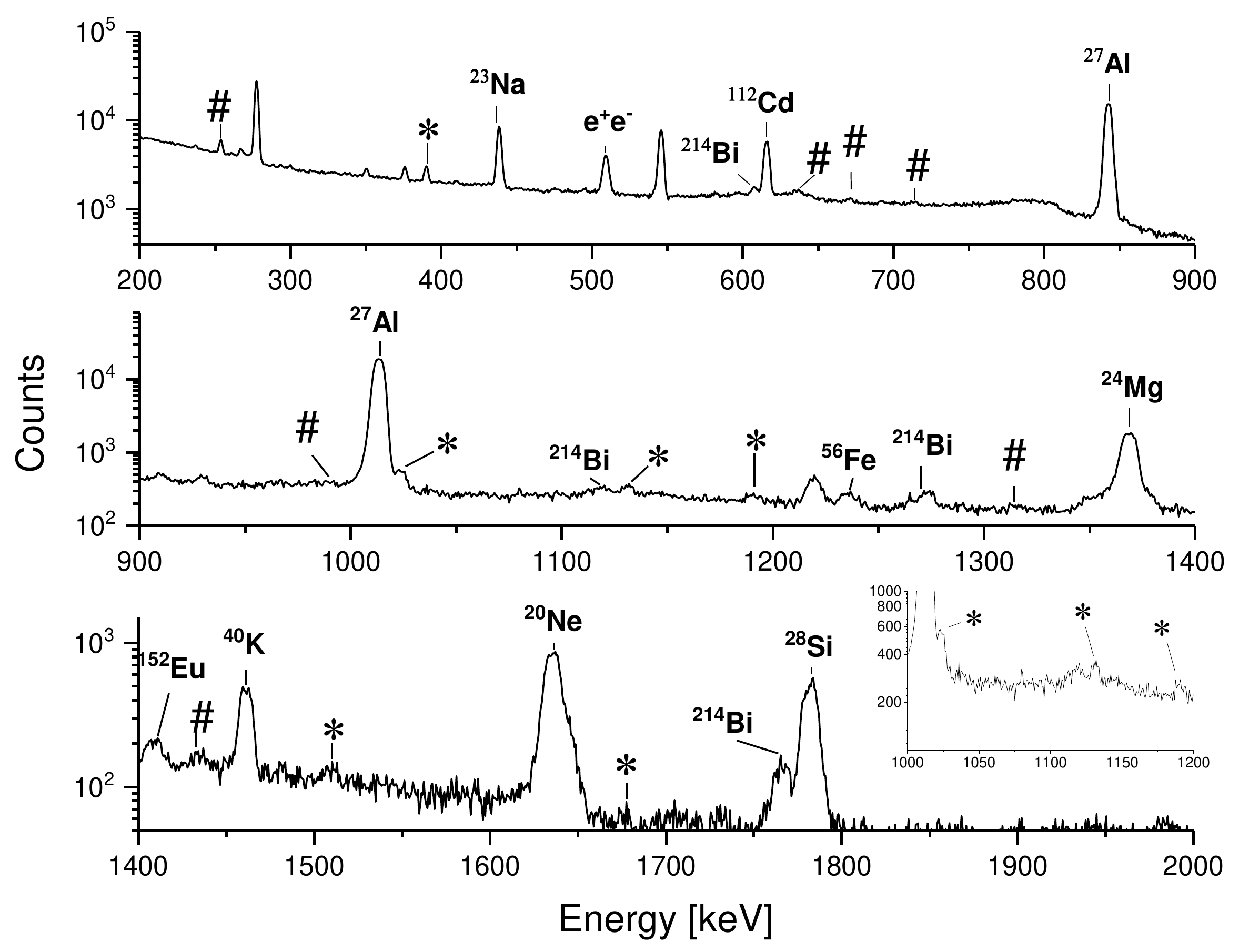}
\caption{\label{fig:112Cd_spectrum}%
Horizontal split--view (0.3--2.0~MeV) of a typical spectrum from the
reaction $\isotope[112]{Cd}+p$, recorded in singles in the detector placed at
90$^{\circ}$, at a beam energy of 3.4~MeV. Photopeaks feeding the g.s. of
\isotope[113]{In} are marked with $\ast$'s, whereas transitions to the isomeric
state of \isotope[113]{In} are denoted with $\#$'s. A more detailed view of the area containing the observed photopeaks for the ground state transition is presented in the inset. Major background lines,
which are usually observed in the present setup, originating from natural
radioactivity (e.g. \isotope[40]{K}, \isotope[214]{Bi}) or elements
present in the beamline components (e.g. \isotope[27]{Al}, \isotope[28]{Si})
are also labeled. Please note that the subfigure y-axes are not in scale.}
\end{figure}

The cross section for the reaction
\isotope[112]{Cd}$(p,\gamma)$\isotope[113]{In}$_{gs}$
can be estimated from the relation~\cite{rolfs}:
\begin{equation}
    \sigma_{gs}=\frac{A}{N_A}\frac{Y}{\delta}
\end{equation}
where $A$ is the atomic mass of the target in atomic mass units (a.m.u.),
$N_A$ is Avogadro's constant, $\delta$ is the actual target thickness, as deduced experimentally by independent dedicated measurements (e.g. RBS for the \isotope[114]{Cd} target), and $Y$ is the absolute yield of the reaction. The latter can be deduced as:
\begin{equation}
    Y=\sum_{i}^n Y_i
    \label{eq:ib_yield_tot}
\end{equation}
where $Y_i$ is the absolute yield of the transition $i$, averaged over angles
$\theta_j$, and calculated as: 
\begin{equation}
 Y_i=\frac{N_i(\theta_j)}{N_p\epsilon_{abs}(\theta_j)}
 \label{eq:ib_yield}
\end{equation}
where, at a measuring angle $\theta_j$, $N_i(\theta_j)$ is the dead-time-corrected
intensity of a photopeak of interest, $\epsilon_{abs}(\theta_j)$ is the absolute
efficiency of the detector, and $N_p$ is the number of incident protons on the
target. From the level scheme of the residual nucleus \isotope[113]{In}, the
following five transitions feeding the ground state were observed in the
in--beam spectra (Fig.~\ref{fig:112Cd_spectrum}), having statistics above
the background (see Ref.~\cite{Psaltis} for a partial level scheme and
Ref.~\cite{nndc} for the data):

\begin{tabular*}{\columnwidth}{rc}\\
$5/2^+_1\rightarrow 9/2^+_{gs}$ & $E_{\gamma}=1024$~keV \\
$5/2^+_2\rightarrow 9/2^+_{gs}$ & $E_{\gamma}=1132$~keV \\
$7/2^+_1\rightarrow 9/2^+_{gs}$ & $E_{\gamma}=1191$~keV \\
($7/2^+,9/2^+)\rightarrow 9/2^+_{gs}$ & $E_{\gamma}=1509$~keV \\
unknown~$\rightarrow 9/2^+_{gs}$ & $E_{\gamma}=1676$~keV \\
\label{112Cd_en}
\end{tabular*}

Results for the ground state cross section are tabulated in Table~\ref{tab:112}
and plotted in Fig.~\ref{fig:112Cd_gs}.

\subsubsection{Activation measurements}
\label{112activ}

The isomeric transition $1/2^-_1 \rightarrow 9/2^+_{gs}$ ($E_{\gamma}=391.7$~keV) 
in \isotope[113]{In} is characterized by a half--life, $t_{1/2}=99.476(23)$ min
\cite{nndc}. Due to the particular lifetime of this transition, the activation
method was employed for the measurement of its absolute yield. It has to be noted that the isomeric transition is associated with a large spin difference ($\Delta I =4$) and a parity flip. An additional
measurement of the cross section of the isomeric state was performed using the
in beam method discussed in~\ref{112ib}.

At each beam energy, the target was irradiated sufficiently long for the process to reach saturation (approximately about 5 half-lives). In our case, the irradiation lasted approximately $4t_{1/2}$. The isomeric cross section was
evaluated using the standard relation~\cite{Psaltis, activ1},
\begin{equation}
    \label{eq:activ}
    \sigma_{is}=\frac{A\lambda e^{\lambda t_w}}{N_t\phi\epsilon_{abs}I_{\gamma}(1-e^{\lambda t_{irr}})(1-e^{\lambda t_c})}
\end{equation}
where $A$ is the number of events under the photopeak corresponding to the
isomeric transition, $I_{\gamma}$ the probability of $\gamma$ ray emission,
$\lambda$ the decay constant of the particular transition, $N_t$ the surface
number density of target nuclei, $\epsilon_{abs}$ the absolute efficiency of
the detector, $\phi$ the incident proton flux during irradiation, and $t_w$,
$t_c$ and $t_{irr}$ are the waiting (or cooling) time of the sample,
the counting time, and the irradiation time of the sample, respectively.
For the case of \isotope[113]{In}, $I_{\gamma}=0.6494(17)$,
$\lambda=116.133(27) \times 10^{-6}$~s$^{-1}$~\cite{ensdf, blachot}.

The results for the isomeric cross sections, deduced with the activation method
are plotted with black upside--down triangles in Fig.~\ref{fig:112Cd_is}. Errors were evaluated
taking into consideration the uncertainties from photopeak integration, the
detector efficiencies, and the charge deposition on the target during the
irradiation of the sample. Cross-section results for the isomeric state deduced
with the in-beam method, by measuring all transitions populating the isomeric
state, are plotted in the same figure (empty circles with black outline). The
results for the isomeric cross sections are tabulated in Table~\ref{tab:112}.
The two datasets corresponding to each energy value, along with the percentage
absolute deviation of the cross sections, as those have been deduced with the
in--beam and activation methods, are listed in Table~\ref{tab:is_ib_activ}. The
maximum value (8\%) was recorded for 3.6~MeV, proving the reliability of using
both methods in combination. It has to be noted that due to their different nature, the two techniques have different systematical uncertainties.

The main contributions to the cross section uncertainty are common for both techniques, and arise from:
\begin{itemize}
    \item The target thickness, with a relative uncertainty of $\sim 5\%$ and $3\%$ for \isotope[112]{Cd} and \isotope[114]{Cd}, respectively.
    \item The detection efficiency, which varies with the energy of the photopeak considered, therefore it is not a fixed value. The maximum relative uncertainty due to the detection efficiency does not exceed $10\%$ in any case.
    \item The uncertainty in the peak area determination, which is the major factor contributing to the final cross section uncertainties. As cited in Tables~\ref{tab:112}, \ref{tab:114} and \ref{tab:is_ib_activ}, the relative cross section uncertainties can reach $\sim 50\%$, depending on the transition examined.
    \item For the activation technique, contributions to the cross section also arise from the uncertainties in the irradiation time $t_{irr}$, counting time $t_c$, and waiting time $t_w$. However, we took special care to minimize these contributions, resulting in relative uncertainties $\ll 1\%$.
\end{itemize}

\subsubsection{Total cross sections and astrophysical $S$ factors}
\label{112cs}

\begin{figure}
 \centering
 \label{fig:112Cd}
 \begin{subfigure}[ht]{0.49\textwidth}
 \centering
 \includegraphics[width=\textwidth]{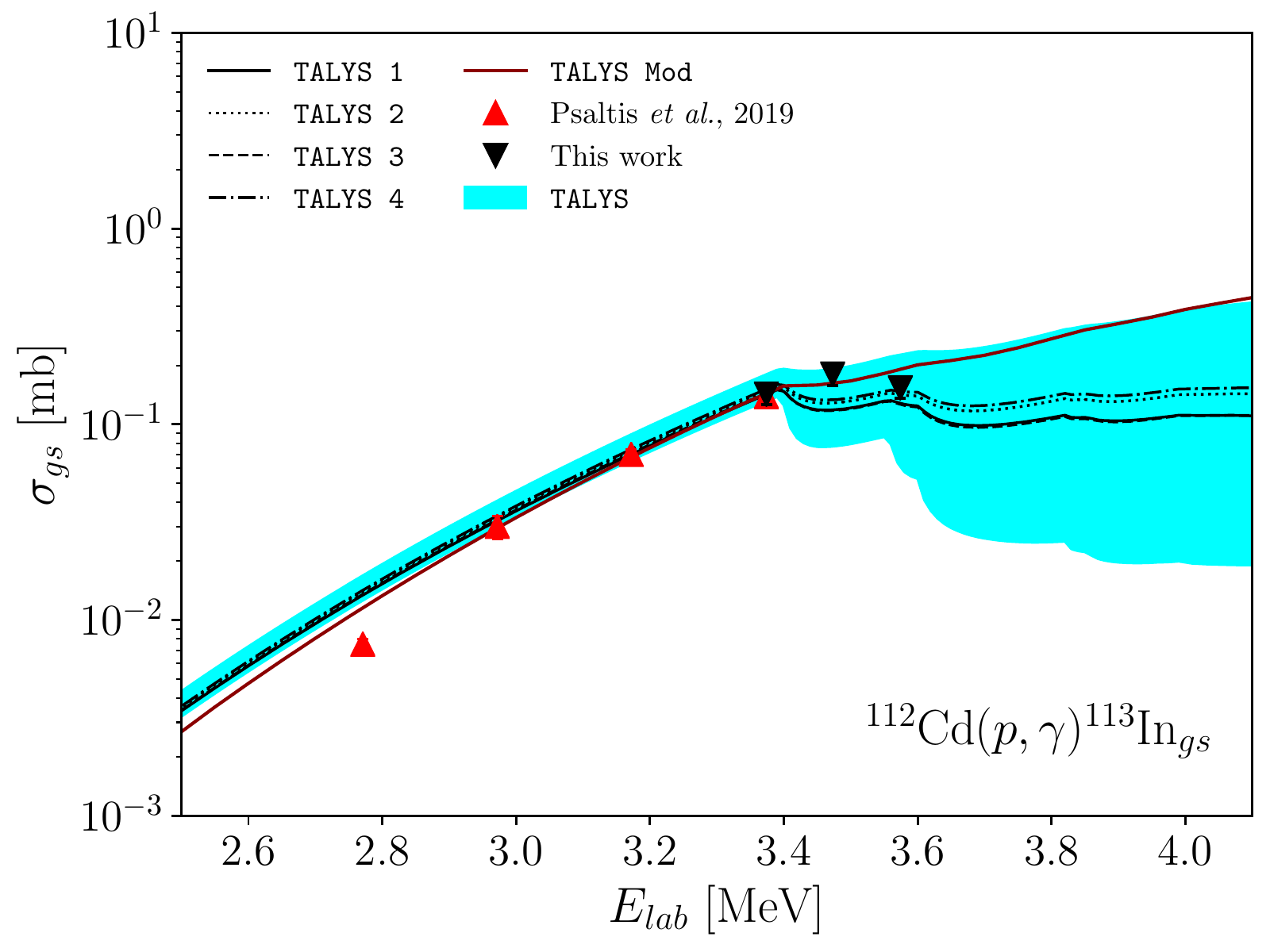}
 \caption{\label{fig:112Cd_gs}%
 Ground--state cross sections for the $(p,\gamma)$ channel deduced with
 the in--beam method. Energies are shown in the laboratory system. The shaded area
 corresponds to the full range of calculated values with every combination of models
 employed. The lines correspond to the best data--matching calculations (see text for
 details).}
 \end{subfigure}
 \hfill
 \begin{subfigure}[ht]{0.49\textwidth}
 \includegraphics[width=\textwidth]{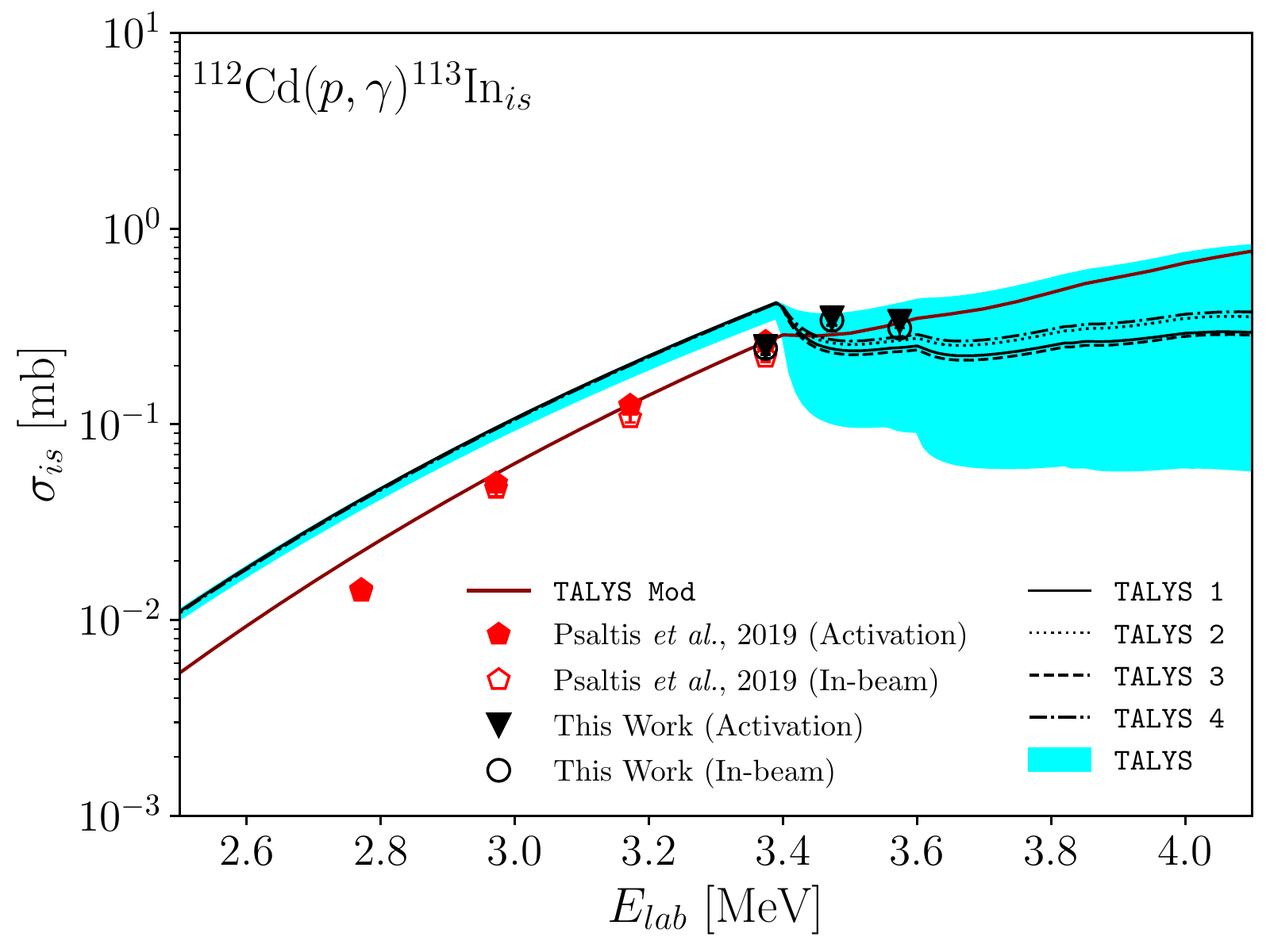}
 \caption{\label{fig:112Cd_is}%
 Measured isomeric cross sections with both the activation (solid black upside--down triangles)
 and the in--beam (empty circles) methods. A good agreement is observed between the results
 of the two methods. The lines and shaded area are as in Fig.~\ref{fig:112Cd_gs}.}
 \end{subfigure}
 \begin{subfigure}[hb]{0.49\textwidth}
 \centering
 \includegraphics[width=\textwidth]{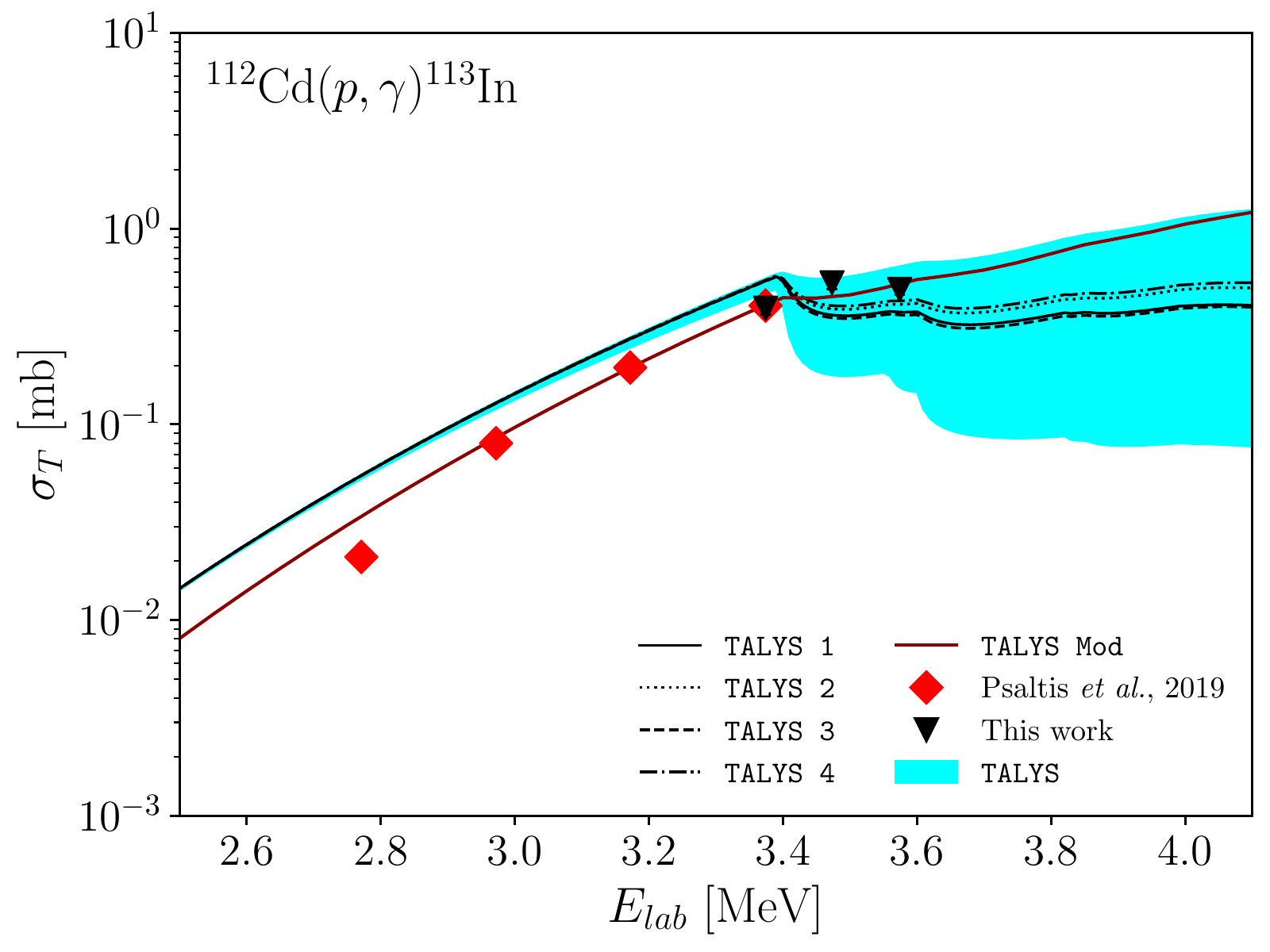}
 \caption{\label{fig:112Cd_tot}%
 As in Fig.~\ref{fig:112Cd_gs} but for the total cross sections of the reaction \isotope[112]{Cd}$(p,\gamma)$\isotope[113]{In}, deduced from the in-beam and activation methods.}
 \end{subfigure}
\hfill
 \begin{subfigure}[hb]{0.49\textwidth}
 \centering
 \includegraphics[width=\textwidth]{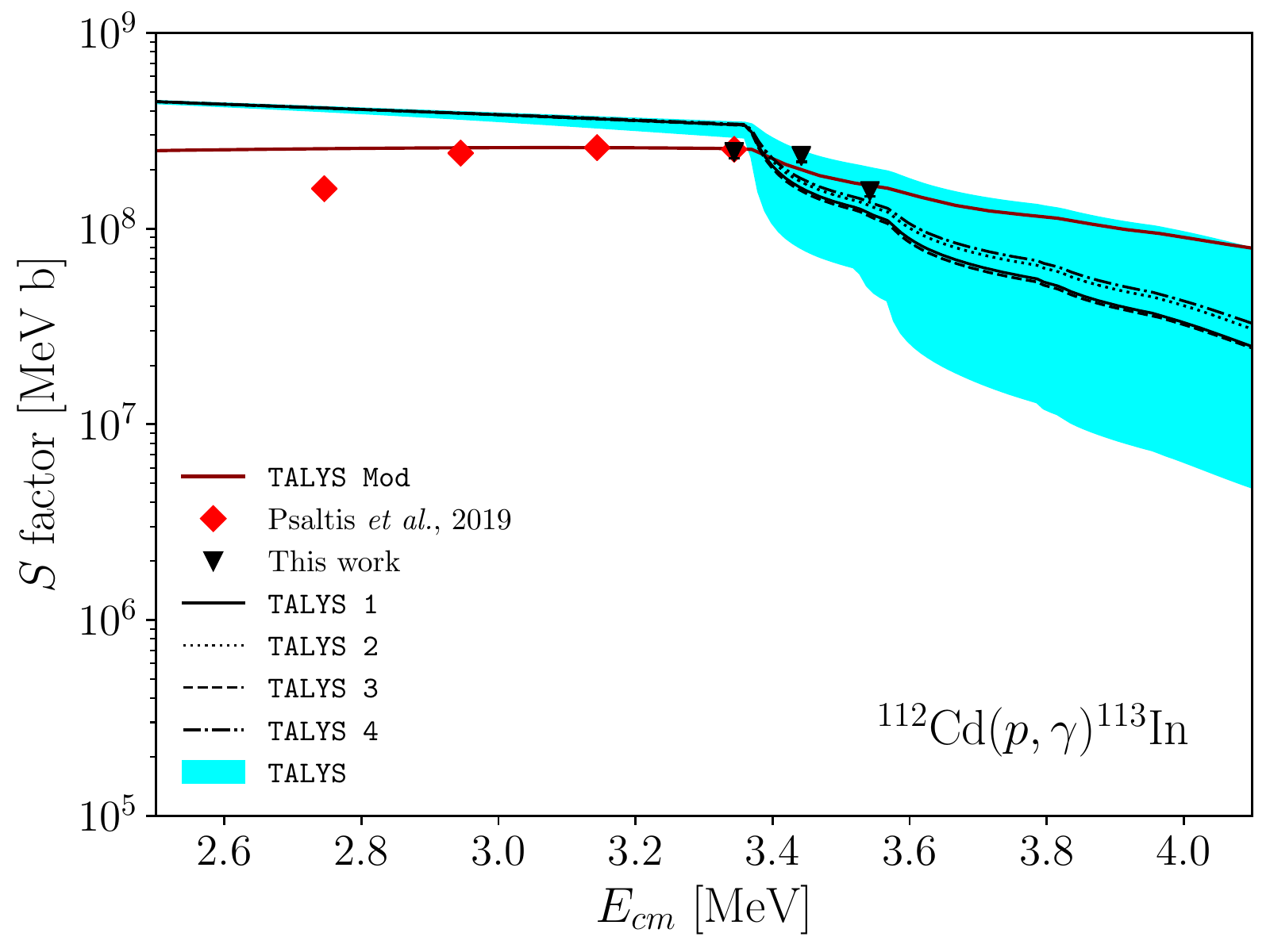}
 \caption{\label{fig:112Cd_sfactor}%
 As in Fig.~\ref{fig:112Cd_tot} but for astrophysical $S$ factors. The difference in this case is that energies are shown in the center-of-mass system.}
 \end{subfigure}
\caption{}
\end{figure}

The total cross sections, $\sigma_T$, of the reaction \isotope[112]{Cd}$(p,\gamma)$\isotope[113]{In}
have been evaluated by adding the cross sections of all transitions feeding
the ground state of \isotope[113]{In} (summing to the in-beam cross-section
$\sigma_{gs}$), and the cross sections of the isomeric state, $\sigma_{is}$,
as deduced with the activation technique (described in~\ref{112activ}), thus
\begin{equation}
    \label{eq:cs_tot}
    \sigma_T=\sigma_{gs}+\sigma_{is}
\end{equation}
The results for the total cross sections for the reaction are tabulated in
Table~\ref{tab:112} and plotted in Fig.~\ref{fig:112Cd_tot}.

After measuring the total cross sections, the astrophysical $S$ factors can be determined through the relation
\begin{equation}
    \label{eq:sf}
    S(E)=E\sigma(E)e^{2\pi\eta}
\end{equation}
where $\eta$ is the Sommerfeld parameter~\cite{sfactor}. The $S$ factor is a quantity
of special importance for astrophysical applications, due to its smooth variation with
energy, as compared to the cross section, allowing for safer extrapolations to experimentally
inaccessible energies, while also serving as a useful quantity for reaction network
calculations~\cite{Psaltis}. The results for the astrophysical $S$ factor are also
tabulated in Table~\ref{tab:112} and plotted in Fig.~\ref{fig:112Cd_sfactor}.
\begin{table}[ht]
\centering
\caption{\label{tab:112}%
Cross sections and astrophysical $S$ factors for the reaction \isotope[112]{Cd}$(p,\gamma)$\isotope[113]{In}.}
\begin{tabular}{ccccccc}\hline\hline
$E$ (lab) & $E_{eff}$ (lab) & $E_{eff}$ (c.m.) & $\sigma_{gs}$ & $\sigma_{is}$ & $\sigma_{T}$ & $S$ factor \\

 [MeV] & [MeV] & [MeV] & [mb] & [mb] & [mb] & [$\times 10^8$~MeV~b] \\
\hline
3.400 & 3.374 & 3.344 & 0.143(17) & 0.250(23) & 0.393(29) & 2.47(18) \\
3.500 & 3.473 & 3.442 & 0.181(23) & 0.35(3) & 0.53(4) & 2.36(17) \\
3.600 & 3.574 & 3.542 & 0.154(18) & 0.336(26) & 0.49(3) & 1.56(10) \\\hline\hline
\end{tabular}
\end{table}

All energies selected for the experiment reside inside the Gamow window for the reaction (see Table~\ref{tab:112} for details).

\subsection{The reaction \isotope[114]{Cd}$(p,\gamma)$\isotope[115]{In}}
\label{ssec:114Cd}

\subsubsection{In-beam measurements}
\label{114ib}

\begin{figure*}[ht]
\centering
\includegraphics[width=0.6\textwidth]{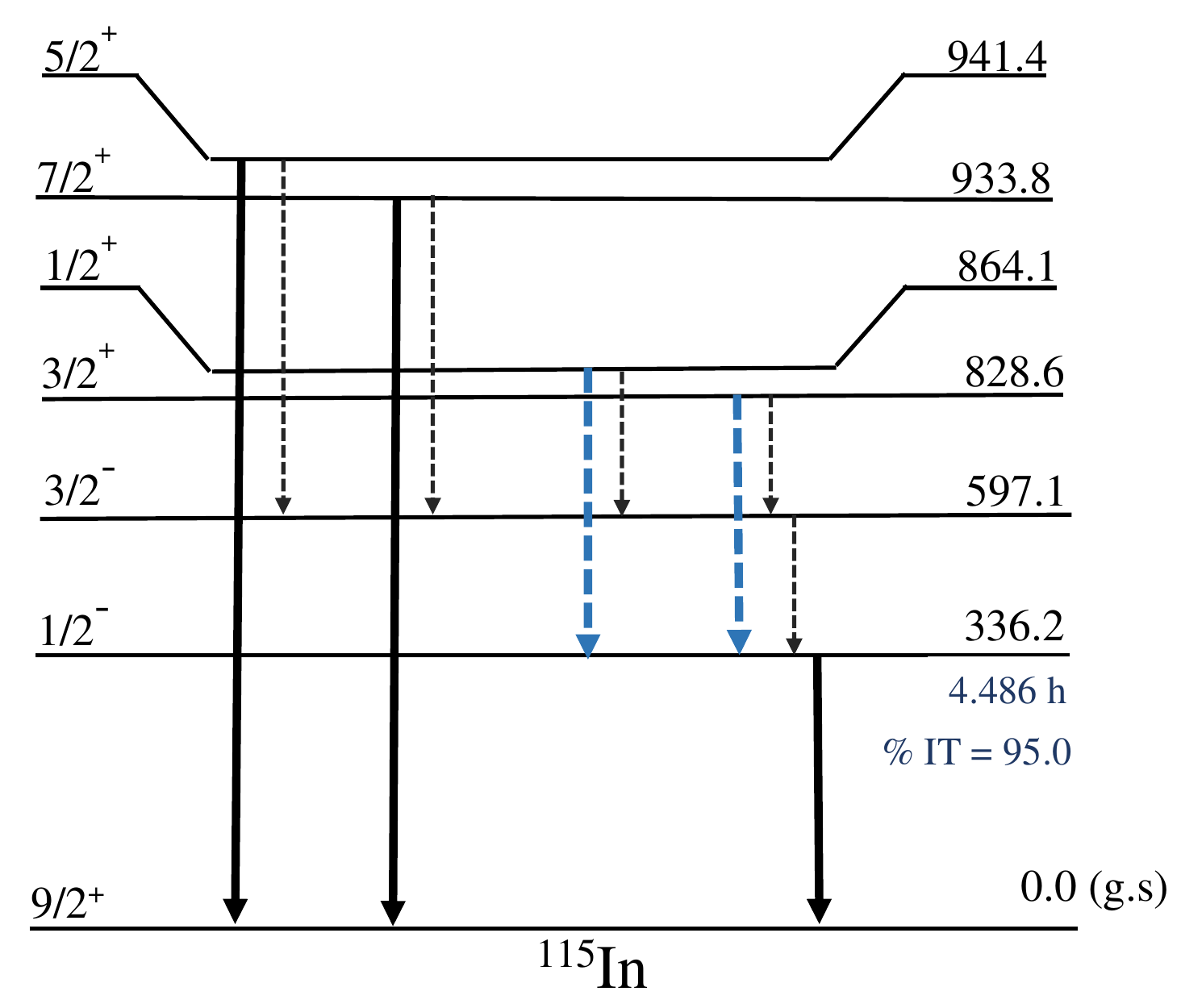}
\caption{\label{fig:115In_level_scheme}%
A partial level scheme of the low-lying energy levels of \isotope[115]{In}, adapted from Ref.~\cite{nndc}. The solid arrows represent decays feeding the ground state of \isotope[115]{In}, and were observed during our measurements. See the transitions denoted with $\ast$'s in Fig.~\ref{fig:114Cd_spectrum}. The thick dashed blue arrows represent transitions feeding the isomeric state, that were observed during our measurements (denoted with \#'s in Fig.~\ref{fig:114Cd_spectrum}) and used for the determination of the isomeric cross section using the in--beam technique. Low--lying transitions, that were not used in the current analysis are denoted with thin dashed black arrows.}
\end{figure*}

\begin{figure}[ht]
\centering
\includegraphics[width=0.9\textwidth]{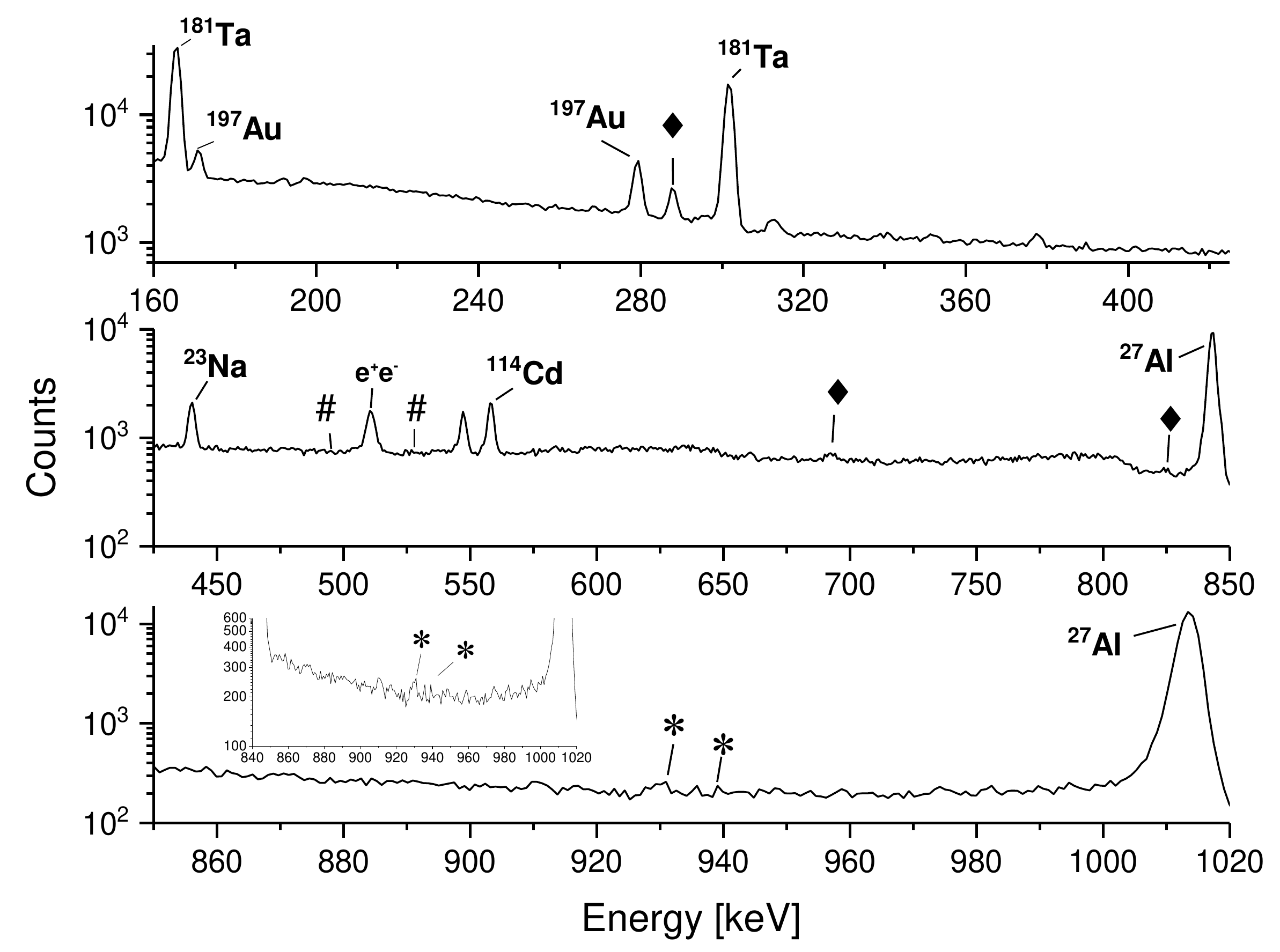}
\caption{\label{fig:114Cd_spectrum}%
Horizontal split-view (0.16--1.02~MeV) of a typical spectrum from the
reaction $\isotope[114]{Cd}+p$, recorded in singles in the detector placed at
55$^{\circ}$, at a beam energy of 4.0~MeV. Photopeaks feeding the g.s. of
\isotope[115]{In} are marked with $\ast$'s, whereas transitions to the isomeric
state of \isotope[115]{In} are denoted with $\#$'s. Transitions to the ground
state of \isotope[114]{In} are denoted with black diamonds. Major background
lines, which are usually observed in the present setup, originating from natural
radioactivity or other processes (e.g. the 511~keV e$^+$e$^-$ annihilation
photopeak) or elements present in the beamline components (e.g. \isotope[27]{Al},
\isotope[nat]{Ta}) are also labeled. A more detailed view of the area containing
the observed photopeaks for the ground state transition is presented in the inset.
(Note: the 941~keV photopeak could only be observed in the summed spectra,
contributing only a very small fraction to the ground state cross section.
However, its location is marked with an asterisk in the inset, for the sake
of completeness). No photopeaks of interest to the studied reactions were
observed beyond the depicted energy range. Please note that the subfigure
y-axes are not in scale.}
\end{figure}

\begin{figure}
\centering
\label{fig:114Cd}
\begin{subfigure}[ht]{0.49\textwidth}
\includegraphics[width=\textwidth]{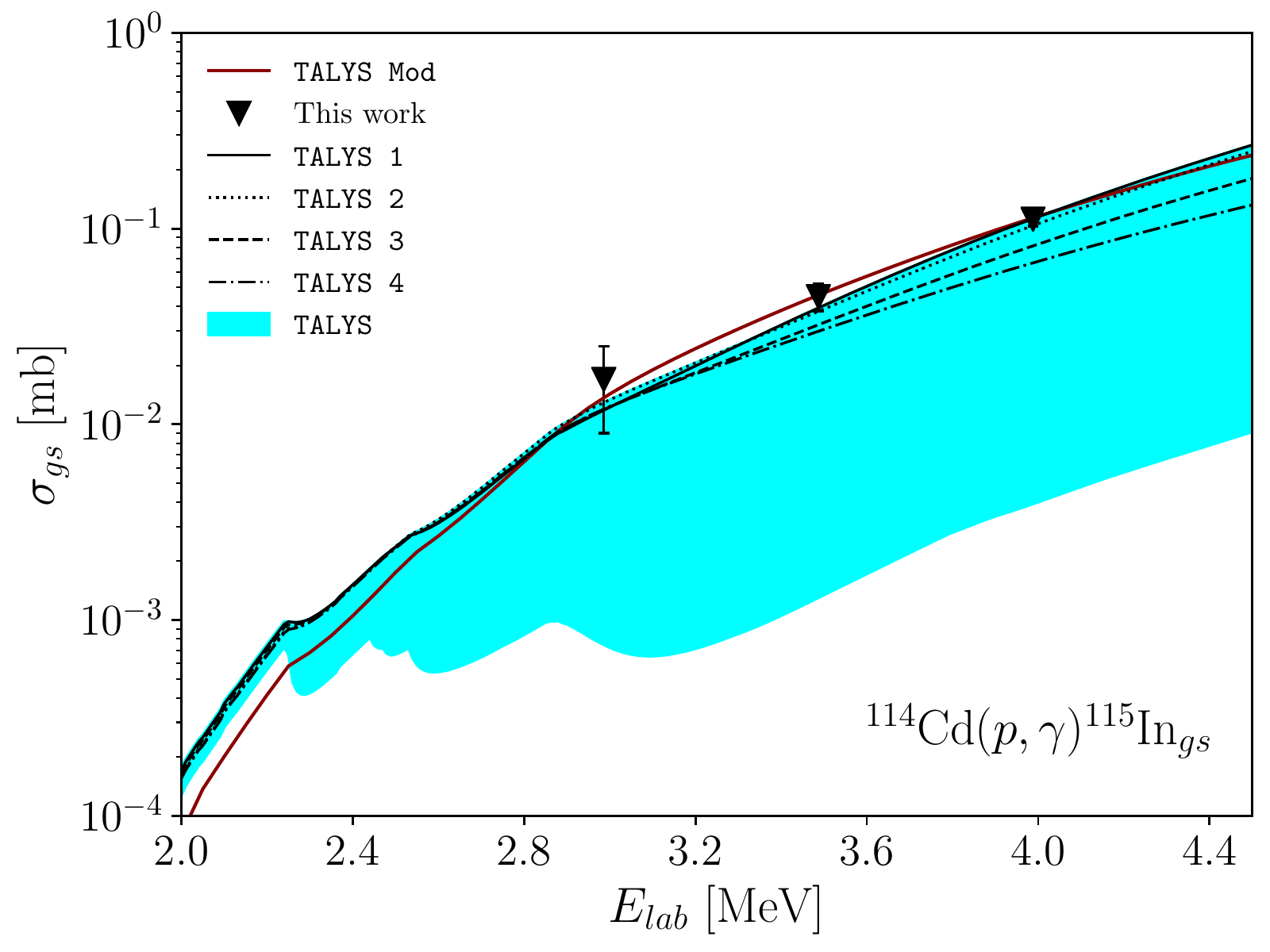}
\caption{\label{fig:114Cd_gs}%
Ground-state cross sections for the $(p,\gamma)$ channel deduced with
the in-beam method. Energies are shown in the laboratory system. The shaded area
corresponds to the full range of calculated values with every combination of models
employed. The lines correspond to the best data-matching calculations (see text for
details).}
\end{subfigure}
\hfill
\begin{subfigure}[ht]{0.49\textwidth}
\centering
\includegraphics[width=\textwidth]{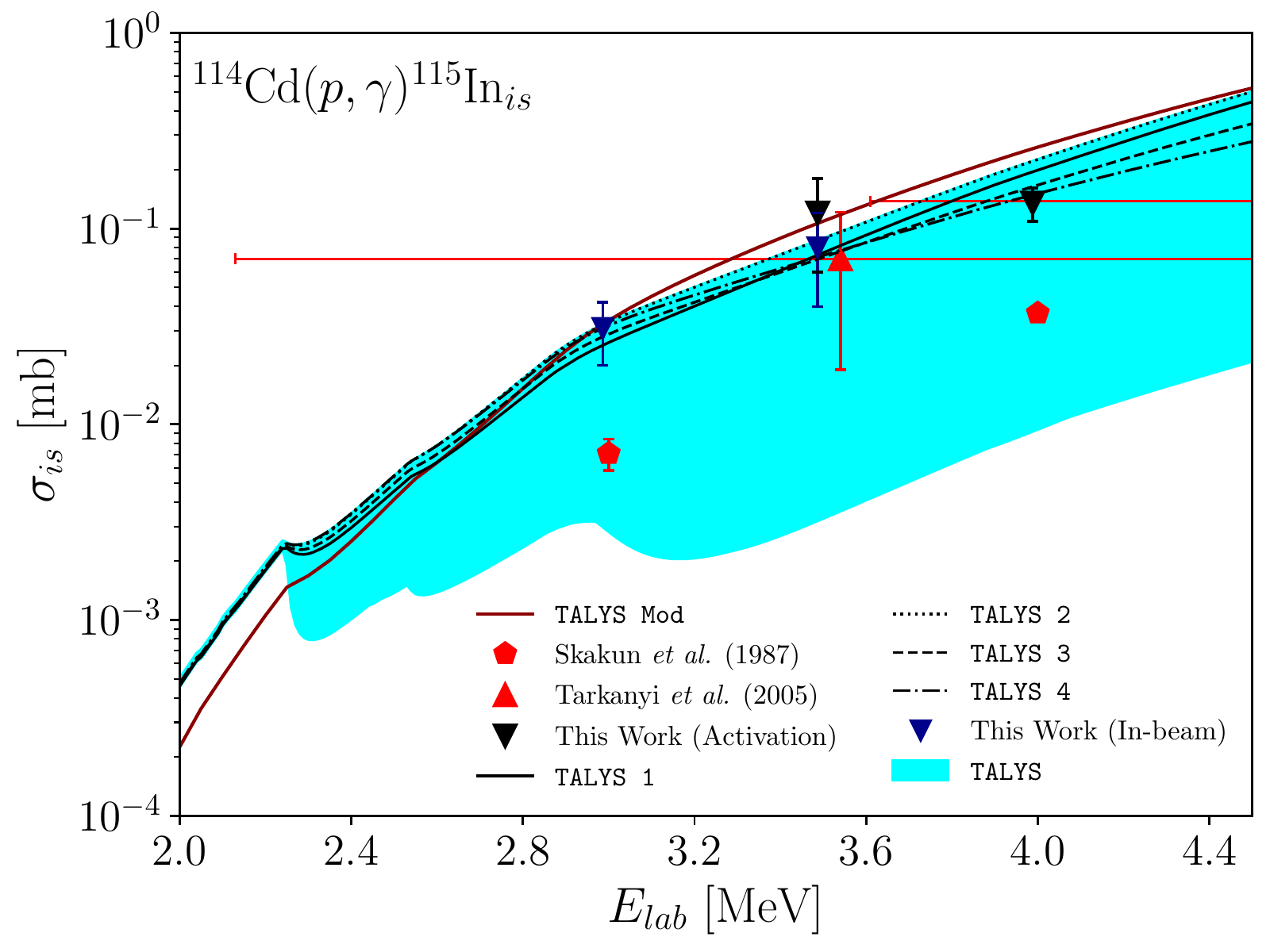}
\caption{\label{fig:114Cd_is}%
Measured isomeric cross sections with both the activation (solid black upside--down triangles) and
the in-beam (dark blue upside--down triangles) methods. The lines and shaded area are as in Fig.~\ref{fig:114Cd_gs}.}
\end{subfigure}
\begin{subfigure}[hb]{0.49\textwidth}
\centering
\includegraphics[width=\textwidth]{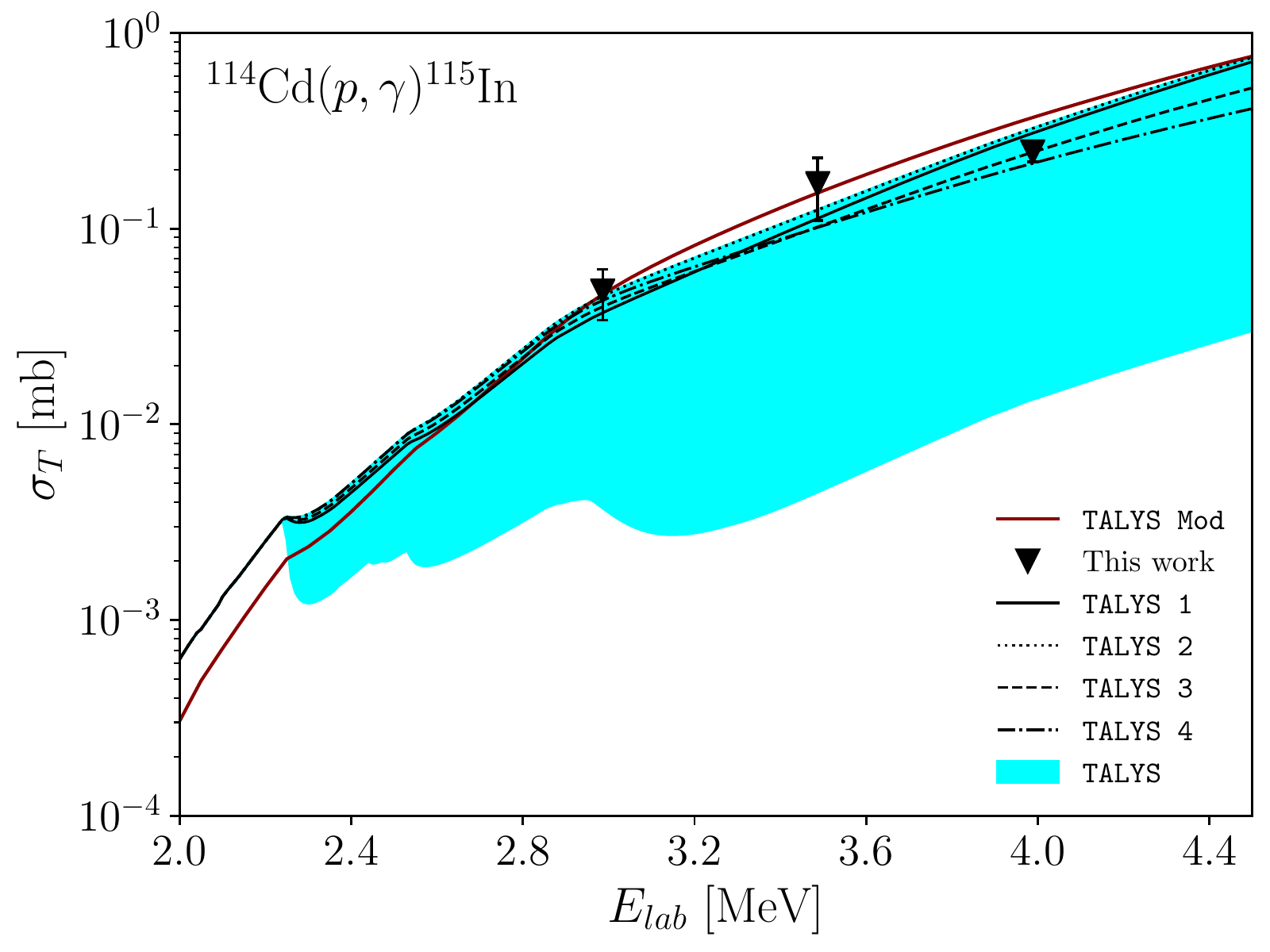}
\caption{\label{fig:114Cd_tot}%
As in Fig.~\ref{fig:114Cd_gs} but for the total cross sections of the reaction \isotope[114]{Cd}$(p,\gamma)$\isotope[115]{In}, deduced from the in-beam and activation methods.}
\end{subfigure}
\hfill
\begin{subfigure}[hb]{0.49\textwidth}
\centering
\includegraphics[width=\textwidth]{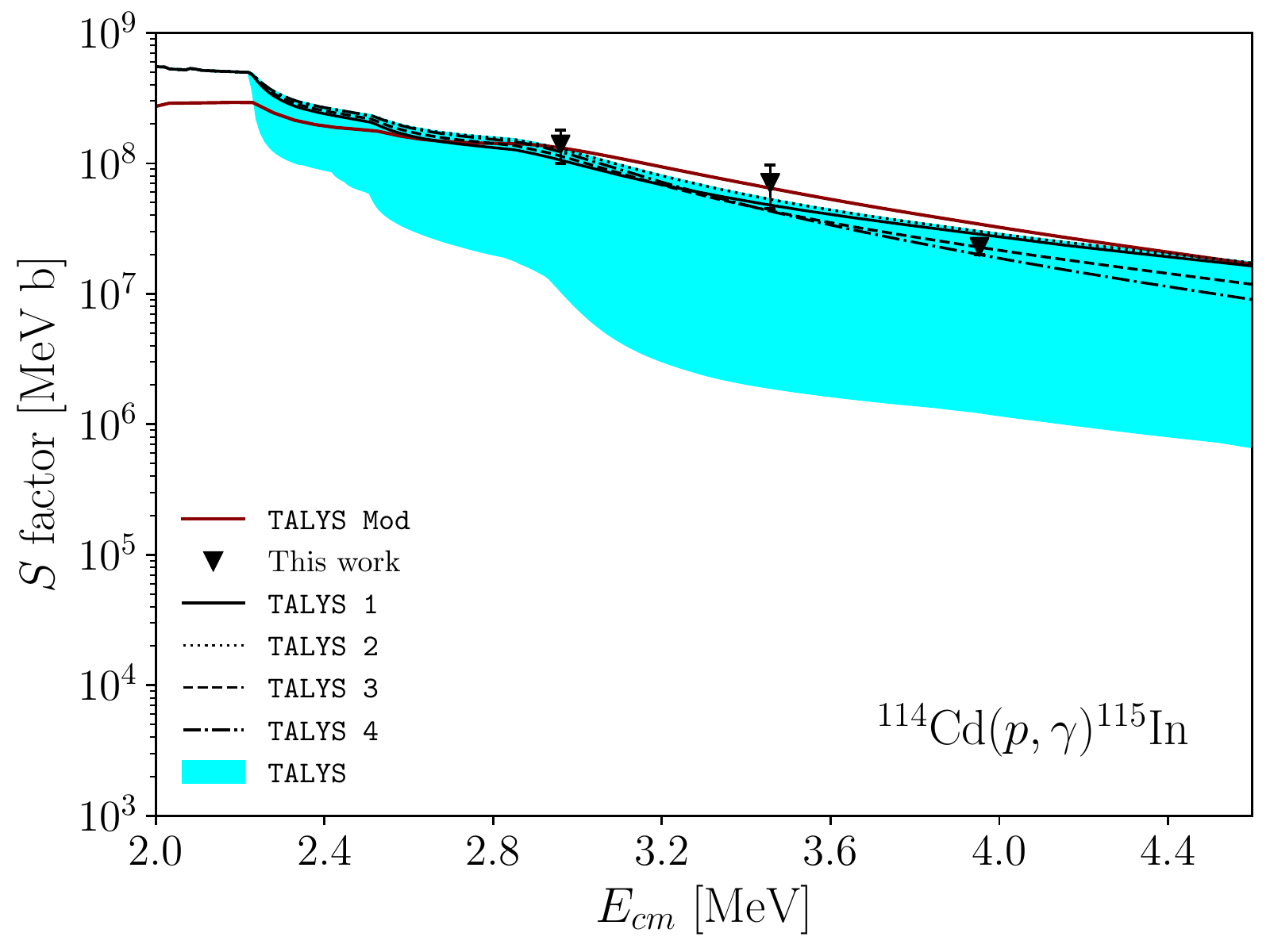}
\caption{\label{fig:114Cd_sfactor}%
As in Fig.~\ref{fig:114Cd_tot} but for astrophysical $S$ factors. The difference in this case is that energies are shown in the center--of--mass system.}
\end{subfigure}
\caption{}
\end{figure}

Following a similar analysis as in the case of the ground state transition in \isotope[113]{In},
the cross section for the reaction \isotope[114]{Cd}$(p,\gamma)$\isotope[115]{In}$_{gs}$ was
evaluated in-beam, with the use of Eqs.~(\ref{eq:ib_yield_tot}) and (\ref{eq:ib_yield}). From
the level scheme of \isotope[115]{In} (see Fig.~\ref{fig:115In_level_scheme} for a partial level
scheme of \isotope[115]{In}, adapted from~\cite{nndc}), two transitions to the ground state of
\isotope[115]{In} were observed in our experiments with statistics above the background (denoted
with asterisks in Fig.~\ref{fig:114Cd_spectrum}):

\begin{tabular*}{\columnwidth}{rc}\\
$7/2^+_1\rightarrow 9/2^+_{gs}$ & $E_{\gamma}=934$~keV \\
$5/2^+_2\rightarrow 9/2^+_{gs}$ & $E_{\gamma}=941$~keV \\
\label{114Cd_pg_en}
\end{tabular*}
The results for the ground state cross sections are tabulated in Table~\ref{tab:114} and plotted in Fig.~\ref{fig:114Cd_gs}.

\subsubsection{Activation measurements}
\label{114activ}

The study of the isomeric $1/2^-_1 \rightarrow 9/2^+_{gs}$ transition in
\isotope[115]{In} presented additional difficulties, compared to the case of
\isotope[113]{In}, due to its particular lifetime of
$t_{1/2}=4.486(4)$~h~[\cite{nndc}].

The \isotope[114]{Cd} target was irradiated for approximately 1.5--2 half-lives,
and thus, the population of the isomeric state in the produced \isotope[115]{In}
was significantly lower compared to the case of \isotope[113]{In}, resulting
in increased statistical uncertainties regarding photopeak integration for the
isomeric transition.

For \isotope[115]{In}, $I_{\gamma}=0.4590(10)$ and $\lambda=42.92(4) \times 10^{-6}$~s$^{-1}$.

\subsubsection{Total cross sections and astrophysical $S$ factors}
\label{114cs}

Using Eq.~(\ref{eq:cs_tot}) the total cross sections for the $(p,\gamma)$ channel
were evaluated. For the isomeric transition at beam energy of 3.0~MeV the value
of $\sigma_{is}$ used in the calculation of $\sigma_T$ was the one resulting
from in-beam measurement (see also Table~\ref{tab:114}). This was due to a
faulty sensor which results in warming up of the HPGe detectors during the
activation measurements, which prevented the extraction of an activation cross 
section value for the 3.0~MeV energy point. Given the very good agreement
between $\sigma_{is}^{in-beam}$ vs $\sigma_{is}^{activ}$ obtained for the rest
of our measurements (see also Sec.~\ref{conclusions}), we consider this choice not
to have a significant impact on the resulting total cross section value, albeit
possibly accompanied by some small missing yield.

Following the calculation of the total cross sections, the astrophysical $S$ factors were evaluated
from Eq.~(\ref{eq:sf}). The results are tabulated in Table~\ref{tab:114} and plotted in
Figs.~\ref{fig:114Cd_tot}, \ref{fig:114Cd_sfactor}.
\begin{table}[ht]
\centering
\caption{\label{tab:114}%
Cross sections and astrophysical $S$ factors for the reaction \isotope[114]{Cd}$(p,\gamma)$\isotope[115]{In}.}
\begin{tabular}{ccccccc} \hline\hline
$E$ (lab) & $E_{eff}$ (lab) & $E_{eff}$ (c.m.) & $\sigma_{gs}$ & $\sigma_{is}$ & $\sigma_{T}$ & $S$ factor \\

 [MeV] & [MeV] & [MeV] & [mb] & [mb] & [mb] & [$\times 10^8$~MeV~b] \\
\hline
3.000 & 2.986 & 2.960 & 0.017(8) & 0.031(11) & 0.048(14) & 1.4(4) \\
3.500 & 3.487 & 3.457 & 0.045(7) & 0.12(6) & 0.17(6) & 0.71(26) \\
4.000 & 3.988 & 3.953 & 0.112(9) & 0.135(26) & 0.247(28) & 0.23(3) \\
\hline\hline
\end{tabular}
\end{table}

All energies selected for the experiment reside inside the Gamow window for the reaction (see Table~\ref{tab:114} for details).

\subsection{The reaction \isotope[114]{Cd}$(p,n)$\isotope[114]{In}}
\label{ssec:114In}

\begin{figure}[ht]
\centering
\includegraphics[width=0.49\textwidth]{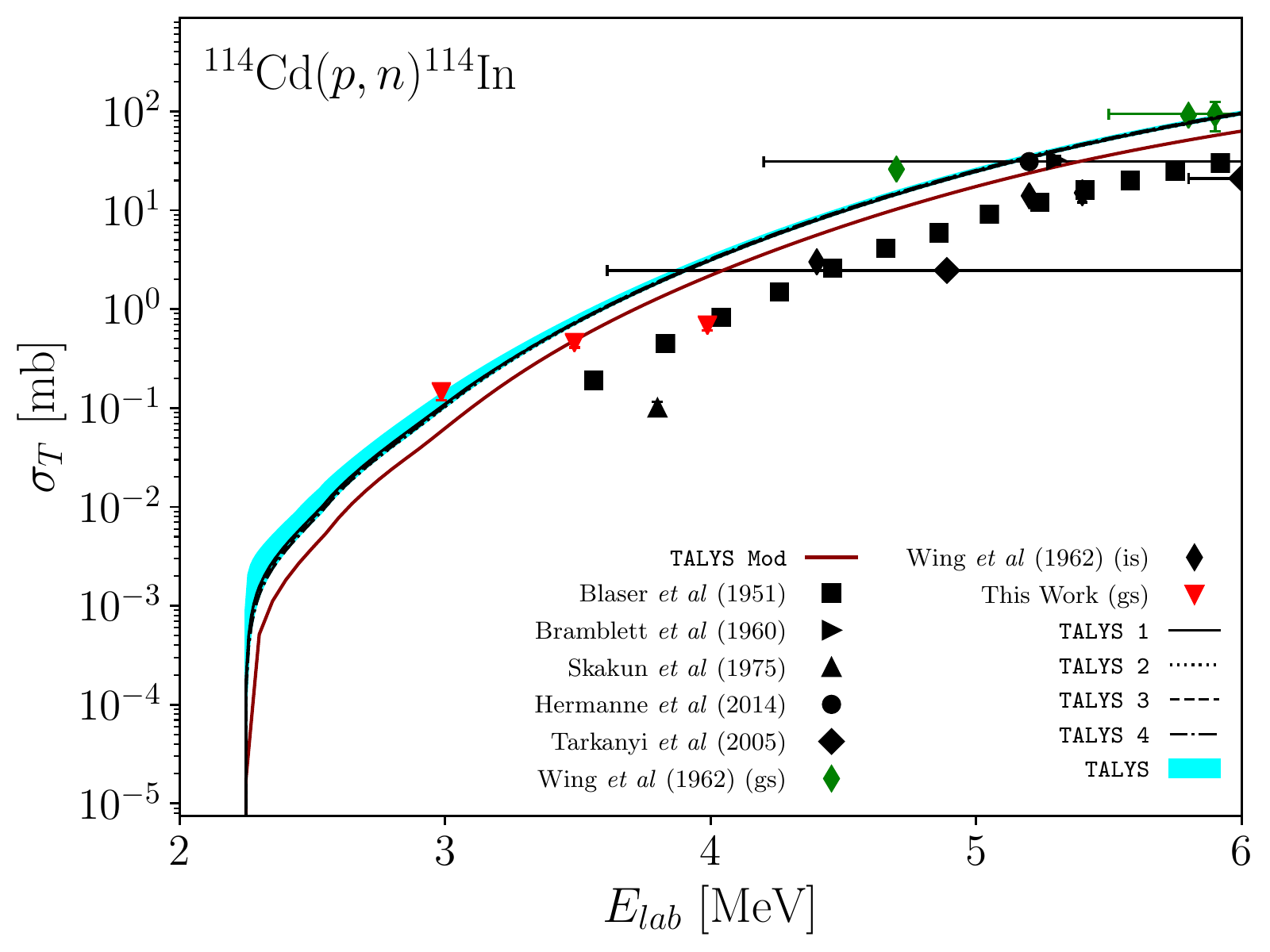}
\caption{\label{fig:114Cd_pn}%
Measured cross sections for the $(p,n)$ channel, deduced with the in-beam method.
The results are compared to \textsc{TALYS} calculations for the total cross sections
of the $(p,n)$ channel, as well as with the existing experimental data, retrieved
from the literature~\cite{Cd4,Cd1,Cd2,Cd3,Cd5,Cd6,Cd9}.}
\end{figure}

The cross section for the reaction \isotope[114]{Cd}$(p,n)$\isotope[114]{In} was determined
by means of the in--beam method, using Eqs.~(\ref{eq:ib_yield_tot}) and (\ref{eq:ib_yield}).
The following transitions to the ground state were observed in the in--beam spectra with statistics
above the background, and thus, were used for the determination of the cross section (see
Ref.~\cite{nndc} for data):

\begin{tabular*}{\columnwidth}{rc}\\
$2^+_1\rightarrow 1^+_{gs}$ & $E_{\gamma}=288$~keV \\
$2^+_1\rightarrow 1^+_{gs}$ & $E_{\gamma}=693$~keV \\
$2^+_2\rightarrow 1^+_{gs}$ & $E_{\gamma}=825$~keV \\
\label{114In_en}
\end{tabular*}
The cross sections for the $(p,n)$ channel in \isotope[114]{In} are tabulated in Table~\ref{tab:114_pn}
and plotted in Fig.~\ref{fig:114Cd_pn}.
\begin{table}
\centering
\caption{\label{tab:114_pn}%
Cross sections for the reaction \isotope[114]{Cd}$(p,n)$\isotope[114]{In}.}
\begin{tabular}{cccc}
\hline\hline
 $E$~(lab) & $E_{eff}$~(lab) & $E_{eff}$~(c.m.) & $\sigma_{gs}$ \\
 
 [MeV] & [MeV] & [MeV] & [mb]\\
\hline
3.000  & 2.986  & 2.960 & 0.146(26) \\
3.500  & 3.487  & 3.457 & 0.46(5) \\
4.000  & 3.988  & 3.953 & 0.69(8) \\ \hline\hline
\end{tabular}
\end{table}

\subsection{Hauser-Feshbach calculations with \textsc{TALYS}}
\label{ssec:talys}

Theoretical calculations using the Hauser-Feshbach statistical model have been performed with
the latest \textsc{TALYS v1.95} code~\cite{talys}. A total of 96 different combinations of the
main components of the model, i.e. the optical potential (OMP) (two default options), the nuclear
level density (NLD) (six default options) and the $\gamma$-ray strength function ($\gamma$SF)
(eight default options) have been used. The models used in the calculations are listed in Table~\ref{tab:talys}. The calculations have bee performed with a 10~keV energy step, from 2--6~MeV.

Both microscopic and phenomenological models have been employed in the calculations, using the
default parameters provided by \textsc{TALYS}. For the OMP, the phenomenological model of
Koning-Delaroche~\cite{omp1} and the semimicroscopic model of Bauge-Delaroche-Girod~\cite{omp2}
have been used. Note that for the studied energy ranges, lying below the respective Coulomb
barriers for each reaction, the OMP, and in particular, its imaginary component, is known to
have a strong energy dependence~\cite{arnould}.

For the NLD, all six available models have been employed in the calculations, namely, the
phenomenological CTM model~\cite{nld1}, the back-shifted Fermi gas model~\cite{nld2}, the
generalized superfluid model~\cite{nld3}, the semi-microscopic level density tables of
Goriely~\cite{nld4} and Goriely et al.~\cite{nld5}, and the time-dependent
Hartree-Fock-Bogolyubov method combined with the Gogny force~\cite{nld6}.

Regarding the $\gamma$SF,
the Kopecky-Uhl~\cite{gsf1} and
Brink-Axel~\cite{gsf2, gsf22} generalized lorentzians were used,
as well as values calculated with the implementation of
the Hartree-Fock-BCS and the
Hartree-Fock-Bogolyubov methods~\cite{gsf3}.
The hybrid model of Goriely~\citep{gsf4}, as well as
Goriely's tables using the TDHFB method~\cite{nld6} were also employed. Last, models using the
temperature-dependent relativistic mean-field method~\cite{gsf5} and the
Hartree-Fock-Bogolyubov method combined with the quasi-random phase approximation using the Gogny D1M
interaction~\cite{gsf6} were implemented.

After performing calculations with all of the possible combinations of the above models,
the maximum and minimum for each energy was determined, thus defining the borders of the light
blue area shown in Figs.~\ref{fig:112Cd_gs}--\ref{fig:112Cd_sfactor}, \ref{fig:114Cd_gs}--\ref{fig:114Cd_sfactor}.

The calculations best describing the experimental data have also been included in the plots
(labeled \textsc{TALYS 1--4}) and are listed in Table~\ref{tab:112_individual_models} for
the reaction \isotope[112]{Cd}$(p,\gamma)$\isotope[113]{In}, and Table~\ref{tab:114_individual_models} for the reaction
\isotope[114]{Cd}$(p,\gamma)$\isotope[115]{In}.

\begin{table}[ht]
\caption{\label{tab:talys}%
Models used for the calculations of the theoretical cross sections with \textsc{TALYS}~\cite{talys}. In total, results from 96 different combinations are presented in this paper.}
\centering
\begin{tabular}{l}\hline\hline
     \textbf{Optical Model Potential (OMP)}  \\\hline
     1.~Koning--Delaroche (KD)~\cite{omp1}\\
     2.~Bauge--Delaroche--Girod (BDG)~\cite{omp2}\\\hline\hline
     \textbf{Nuclear Level Density (NLD)}\\\hline
     1.~Constant--temperature model (CTM)~\cite{nld1}\\
     2.~Back--shifted Fermi gas model (BSFG)~\cite{nld2}\\
     3.~Generalized superfluid model (GSM)~\cite{nld3}\\
     4.~Goriely of Goriely et al.~\cite{nld4}\\
     5.~Tables of Goriely et al.~\cite{nld5}\\
     6.~T--dependent HFB, Gogny force (TDHFB)~\cite{nld6}\\\hline\hline
     \textbf{$\gamma$ Strength Function ($\gamma$SF)}\\\hline
     1.~Kopecky--Uhl~\cite{gsf1}\\
     2.~Brink~\cite{gsf2} and Axel~\cite{gsf22}\\
     3.~Hartree--Fock BCS (HFBCS)~\cite{gsf3}\\
     4.~Hartree--Fock--Bogolyubov (HFB)~\cite{gsf3}\\
     5.~Goriely's hybrid model~\cite{gsf4}\\
     6.~Goriely TDHFB~\cite{gsf5}\\
     7.~T--dependent relativistic mean field (RMF)~\cite{gsf6}\\
     8.~Gogny D1M HFB$+$quasi-random-phase approximation (QRPA)~\cite{gsf6}\\\hline\hline
\end{tabular}
\end{table}

\begin{table}[ht]
\centering
\caption{\label{tab:112_individual_models}%
\textsc{TALYS} combinations that best describe the experimental data for the reaction \isotope[112]{Cd}$(p,\gamma)$\isotope[113]{In} (for abbreviations, see Table~\ref{tab:talys}).}
\begin{tabular}{cccc} \hline\hline
 Model & OMP & NLD & $\gamma$SF \\
\hline
\texttt{TALYS 1}  & KD  & BSFG  & Goriely TDHFB \\
\texttt{TALYS 2}  & KD  & GSM   & Brink--Axel \\
\texttt{TALYS 3}  & KD  & GSM   & Goriely's hybrid \\
\texttt{TALYS 4}  & KD  & TDHFB & Goriely TDHFB \\ \hline\hline
\end{tabular}
\end{table}

\begin{table}[ht]
\centering
\caption{\label{tab:114_individual_models}%
\textsc{TALYS} combinations that best describe the experimental data for the reaction \isotope[114]{Cd}$(p,\gamma)$\isotope[115]{In} (for abbreviations, see Table~\ref{tab:talys}).}
\begin{tabular}{cccc} \hline\hline
 Model & OMP & NLD & $\gamma$SF \\
\hline
\texttt{TALYS 1}  & KD  & Goriely of Goriely et al. & HFB \\
\texttt{TALYS 2}  & KD  & Goriely of Goriely et al. & Goriely's hybrid \\
\texttt{TALYS 3}  & KD  & Goriely of Goriely et al. & Brink-Axel \\
\texttt{TALYS 4}  & KD  & GSM & Goriely TDHFB\\ \hline\hline
\end{tabular}
\end{table}

\begin{table}[ht]
\centering
\caption{\label{tab:talys_mod}%
The parameters employed in the model ``\textsc{TALYS} Mod'', which lead to a simultaneous good description of the experimental combinations that best describe the experimental data for every reaction channel studied in the present paper (for abbreviations, see Table~\ref{tab:talys}). For the OMP, NLD and $\gamma$SF parameters not listed in the table, the default parameter values were adopted. A detailed description of each parameter can be found in the \textsc{TALYS} v1.95 manual~\cite{talys}.}
\begin{tabular}{cccc} \hline\hline
Model  & Adjusted   & \textsc{TALYS}  & Value\\
(\textsc{TALYS} input) & Parameter & input & \\\hline
KD OMP (jlmomp n)  & $v_1$~[MeV]  & v1adjust p& 0.5 \\
        & $r_C$~[fm]  & rcadjust p& 2.1 \\
        & $a_v$~[fm]  & avadjust p& 0.74\\
        & $a_D^w$~[fm]& awdadjust p& 0.5 \\
        &&&\\
GSM NLD (ldmodel 3) & $f_{sc} $ & Rspincut & 0.1 \\
        &&&\\
Brink-Axel $\gamma$SF (strength 2) & Default & Default & Default\\
\hline\hline
\end{tabular}
\end{table}

\begin{table}[ht]
\centering
\caption{\label{tab:is_ib_activ}%
Isomeric cross sections for the reaction $^{112}\mathrm{Cd}(p,\gamma)^{113}\mathrm{In}$ deduced from the activation and the in-beam measurements for the three beam energies (laboratory). In the far right column, the percentage absolute differences between the in-beam results and the activation results are shown. The data sets are plotted in Fig.~\ref{fig:112Cd_is}.}
\begin{tabular}{cccc}\hline\hline
 $E_{eff}$ (lab)& $\sigma_{is}$ (activation) & $\sigma_{is}$ (in-beam) & Deviation \\
 
 [MeV] & [mb] & [mb] & [\%] \\
\hline
3.374  & 0.250(23)  & 0.244(26) & 2 \\
3.473  & 0.35(3)    & 0.34(4)   & 3  \\
3.574  & 0.336(26)  & 0.31(3)   & 8  \\ \hline\hline
\end{tabular}
\end{table}

\label{sec:exp}

\section{Discussion and Conclusions}

An experimental attempt to measure the total reaction cross sections and the $S$ factors
for the \isotope[114]{Cd}$(p,\gamma)$\isotope[115]{In} reaction has been carried out for
the first time, inside the astrophysically important energy regime (at beam energies of
3.0, 3.5 and 4.0~MeV). An extension of a recent experimental effort in \isotope[112]{Cd}
to higher energies than before (3.4, 3.5 and 3.6 MeV) has additionally been carried out.
Both reactions exhibit very low cross sections, in the order of tens to hundreds of $\mu$b
in the energy range studied in the present work. An additional level of complexity in
the present work has been related to the lifetimes of the isomeric states being present
in both nuclei, making the use of two methods to obtain the experimental results imperative.

For the case of populated \isotope[115]{In}, the in--beam $\gamma$ spectroscopy focused on
all prompt $\gamma$ transitions feeding its ground state directly. All visible
transitions in the spectra populating the isomeric state were also included in the calculation
of the cross section. The activation technique was additionally implemented~\cite{activ1, rolfs},
in order to fully account for the contribution of the significantly longer--lived isomeric
state in \isotope[115]{In}.

A general setback of the in-beam method, is its high dependency on the detection limit of
the experimental setup. It has been observed~\cite{Harissopulos, Khaliel, Psaltis} that with
this particular technique some weak transitions may be absent from the spectra, resulting in
some missing strengths, especially when singles--mode is involved. 
Regarding the isomeric transition, a comparison between the in--beam and activation cross
sections shows that no major contributions to the cross section arise due to the missing
higher level transitions. Given the fact that the cross sections reported in this work are
very small (of the order of tens to hundreds of microbarns) we consider the last argument
to hold true also for the ground state cross sections, with the more strongly observed
transitions in the in--beam spectra carrying the largest fraction of the cross section,
while higher level transitions, which do not show up in the spectra, amount to negligible
contributions to the corresponding cross section.

An alternative experimental approach to remedy all that could possibly be the application 
of the 4$\pi$ detection method. Implementation of this method reduces the data analysis to
that of the produced single summing peak, instead of a complex $\gamma$--ray spectrum. The
aforementioned method has been applied successfully for studies in reactions relevant to the 
{\em p}--process~\cite{Spyrou,Foteinou} despite its own constraints, such as the summing peak
detection efficiency, which depends on the $\gamma$--decay scheme.

The results for the isomeric state populated during the reaction 
\isotope[114]{Cd}$(p,\gamma)$\isotope[115]{In} are compared to existing experimental datasets
retrieved from the literature. A better agreement exists with the published dataset by
T\'ark\'anyi et al.~\cite{Cd4}, rather than that by Skakun et al.~\cite{Cd7}. It has to be
noted that large uncertainties in the energy exist in the former, possibly due to the
stack--foil technique employed, on which the current work seems to improve significantly.
None of the earlier works report on experimental data regarding the ground state cross sections for the radiative $(p,\gamma$)
reaction with \isotope[114]{Cd}, thus making the present dataset on the ground state and
total reaction cross sections the first one to be reported for energies inside the Gamow window.

In addition, this work further extends the data for the \isotope[112]{Cd}$(p,\gamma)$\isotope[113]{In}
reaction at energies slightly above the $(p,n)$ energy threshold, but still inside the Gamow
window. A measurement at proton beam energy of 3.4~MeV, which was also performed in the
previous work~\cite{Psaltis}, was revisited in this experiment, aiming to confirm the validity
of our experimental methodology in an independent measurement and analysis, before moving
to higher energies or another nucleus. The present results at this energy show excellent
agreement with the previously published values, as illustrated in
Figs.~\ref{fig:112Cd_gs}-\ref{fig:112Cd_sfactor}, without any normalization considered.
Two more energy points have been measured at energies 3.5 and 3.6~MeV, respectively. As it
is evident in Figs.~\ref{fig:112Cd_gs}-\ref{fig:112Cd_sfactor}, the existence of the neutron
threshold has a clear effect --as expected-- on the trend of the cross sections. While,
below the threshold, the cross sections seem to follow a smooth trend affected largely by
the OMP, the wider phase space above the threshold shows the important role of NLD at this
energy regime. The total reaction cross sections and the corresponding $S$ factors show a
clear deviation from the trend of the data at lower energies. As the neutron threshold for
the \isotope[114]{Cd}$(p,n)$\isotope[114]{In} reaction is at lower energies (2.247~MeV)~\cite{nndc},
no observation similar to the case of \isotope[112]{Cd} is observed; rather a smooth,
monotonous trend is exhibited in the studied energy range (see
Figs.~\ref{fig:114Cd_gs}-\ref{fig:114Cd_sfactor}).

In order to gain further insight on the OMP, NLD and $\gamma$SF parameters involved in the
theoretical calculations, the \isotope[114]{Cd}$(p,n)$\isotope[114]{In} reaction was additionally studied during the same experiment. The deduced experimental cross sections for this reaction channel are shown in Table~\ref{tab:114_pn}, and plotted
with red upside--down triangles in Fig.~\ref{fig:114Cd_pn}. 
At this point, we should note that our measurements are accompanied by some missing yield, resulting from the long--lived isomer in \isotope[114]{In} ($t_{1/2} \approx 49.5$~d \cite{nndc}), and the partial cross section pertaining to the formation of the daughter \isotope[114]{In} nucleus in its ground state (i.e. without the involvement of $\gamma$--ray emission). However, this fact does not affect our conclusions, which are corroborated by comparison of the results with the datasets existing in the literature (Refs.~\cite{Cd1,Cd2,Cd3,Cd4,Cd5,Cd6,Cd9}), along with those pertaining to the corresponding $(p,n)$ channel for \isotope[112]{Cd}.
The results seem to follow the general trend formed by earlier datasets retrieved from the literature,
despite there is a divergence between the absolute experimental data and the theoretical
calculations in the  energy range below $5.0$--$6.0$~MeV, down to the $(p,n)$ energy threshold.
A similar behavior is observed for the corresponding $(p,n)$ channel in \isotope[112]{Cd},
which could possibly be attributed to the fact that the incorporated phenomenological and
semi--microscopic OMPs have been optimized at a significantly higher energy range than the
one the present paper focuses on. {\sc TALYS 1--4} combinations included in Table~\ref{tab:114_individual_models}, which best describe the ground--state cross sections
of the \isotope[114]{Cd}$(p,\gamma)$ channel, seem to also best describe the experimental
$(p,n)$ data reported in this work, along with the data obtained by Hermanne et al.
(black circles in Fig.~\ref{fig:114Cd_pn}) and Wing {\em et al.} (g.s.) (green triangles in
Fig.~\ref{fig:114Cd_pn}). However, the choice of this default set of parameters overestimates
the \isotope[113]{In} isomeric state cross sections significantly, as was also reported
earlier~\cite{Psaltis}.

Based on the new experimental data reported in this paper for every $(p,\gamma)$ channel studied, as well as the respective data sets from the previous work from our group~\cite{Psaltis}, a systematic attempt to determine
the best set of parameters of OMP+NLD+$\gamma$SF entering the {\sc TALYS} theoretical models
was undertaken~\cite{talys_manual}. The goal was to achieve a good description of the experimental
data for each reaction channel studied in a {\em simultaneous} fashion. Due to the very large
number of parameters tested in this step, the full results from the parameter sensitivity
analysis are not included in here; rather the predictions with the ``best--fit'' values are
presented, exclusively, i.e. those providing the best simultaneous fit to all experimental
data. It has to be stressed that these parameters have been used for both reactions without
any further modifications.

Initially, every parameter of the KD OMP~\cite{omp1} was individually tested, by modifying its 
corresponding {\sc TALYS} input variable in small steps with respect to the default value,
and observing the resulting effect on the theoretical predictions for the ground state and
isomeric cross sections in both scale and trend, for the full energy range examined in this
work. Having extracted the ``best--fit'' values for all OMP parameters, the same procedure
was followed for the NLD model parameters, finally arriving at the values tabulated
in Table~\ref{tab:talys_mod}. The Brink--Axel generalized Lorentzian~\cite{gsf2,gsf22}, with its
default parameter values already led to a good description of the experimental data, and was
chosen as the $\gamma$SF of choice throughout.

For the KD OMP, the modified parameters are:
the coefficient of the real component of the volume--central potential of the
energy dependent well depth $V_V(E)$, $v_1$;
the diffuseness parameter of the real and imaginary components of the volume--central
potential of the energy--independent radial part, $a_v$;
the diffuseness parameter of the imaginary component of the surface--central potential
of the energy--independent radial part, $a_D^w$; and 
the Coulomb radius constant of the energy--independent radial part of the Coulomb term,
$r_C$. These parameters have been adjusted by constant multiplication factors, set via
the corresponding \textsc{TALYS} input variables listed in Table~\ref{tab:talys_mod}.
For the GSM NLD,
the spin cut--off, $\sigma^2$, of the nuclear level density model was multiplied by a
constant value, set through the {\sc TALYS} input variable ``Rspincut'' (we have denoted this
parameter with $f_{sc}$ in Table~\ref{tab:talys_mod}). All other parameters, were used
with their default values.

The predictions of the final theoretical {\sc TALYS} model arising from the use of the
modified parameters, as listed in~\ref{tab:talys_mod} is plotted for each reaction channel
studied in the present paper. The corresponding dark red solid line, labelled ``{\sc TALYS}
Mod'' in these plots, leads to a significantly improved agreement for both reaction channels
in a simultaneous fashion. Especially for the isomeric state, the improvement over any of
the default combinations is self--evident.

In conclusion, the set of experimentally deduced cross sections and astrophysical {\em S}
factors reported in the present paper for the proton--induced reactions with \isotope[112,114]{Cd}
provide new information that can support the improvement of reaction network calculations
around the studied mass region. It is certainly necessary to stress that the tailor--made
model described above can not be considered a global description for the nuclei under scrutiny.
As the experimental data are located in a limited energy range, the validity of the tailored
description can only be considered safe inside it. However, the success in achieving a good
simultaneous fit for both ground and isomeric states in \isotope[113,115]{In} produced in the
studied reactions lays the ground for further investigation. Both experimental and theoretical
studies are required to acquire firm insight at the driving mechanisms behind the {\em p}--process
nucleosynthesis and restrict the parameters of the theoretical models in an energy region
where a scarcity of experimental data, even for stable nuclei, still persists.

\label{conclusions}

\section*{Acknowledgements}
We are grateful to M.~Andrianis, A.~Laoutaris, and S.~Nanos for providing beams
during the experiments, and Dr. E.~Ntemou for technical assistance with the target
measurement.
\hfill \\
\begin{minipage}[]{0.65\columnwidth}
AZ and AC acknowledge support
by the Hellenic Foundation for Research and Innovation (HFRI)
and the General Secretariat for Research and Technology (GSRT)
under the HFRI PhD Fellowship grant
(GA. No. 101742/2019 and 74117/2017, respectively).
\end{minipage}
\hfill
\begin{minipage}[r]{0.3\columnwidth}
\hspace{-5mm}\includegraphics[width=30mm]{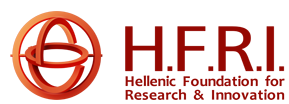}
\end{minipage}

 \bibliographystyle{elsarticle-num} 
 \bibliography{Cd}





\end{document}